\documentclass[nonacm=true]{acmart}
\setcopyright{none}
\renewcommand\footnotetextcopyrightpermission[1]{} 
\makeatletter
\let\@authorsaddresses\@empty
\makeatother

\usepackage[utf8]{inputenc}
\usepackage{hyperref}
\usepackage{enumitem}
\usepackage{dirtytalk}
\usepackage{graphicx}
\usepackage{wrapfig}
\usepackage{float}
\usepackage[multiple]{footmisc}
\usepackage{cleveref}
\usepackage{gensymb}

\AtBeginDocument{%
  \providecommand\BibTeX{{%
    \normalfont B\kern-0.5em{\scshape i\kern-0.25em b}\kern-0.8em\TeX}}}

\begin{document}

\title[Eye Tracking in IR]{Considerations for Eye Tracking Experiments in Information Retrieval}

\author{Michael Segundo Ortiz}
\affiliation{%
 \institution{Carolina Health Informatics Program, University of North Carolina at Chapel Hill}
 \city{Chapel Hill}
 \state{NC}
 \country{USA}}
\email{msortiz@unc.edu}

\renewcommand{\shortauthors}{Ortiz}

\begin{abstract}
In this survey I discuss ophthalmic neurophysiology and the experimental considerations that must be made to reduce possible noise in an eye-tracking data stream. I also review the history, experiments, technological benefits and limitations of eye-tracking within the information retrieval field. The concepts of aware and adaptive user interfaces are also explored that humbly make an attempt to synthesize work from the fields of industrial engineering and psychophysiology with information retrieval.
\end{abstract}

\maketitle
\thispagestyle{empty}

\section{Introduction}
On the nature of learning, I think about my son. A 1-year old at the time of this writing, he plays by waving his arms, looking around, yelling, and putting his mouth on every object in sight. In these moments, I observe him without explicit instruction unless of course danger lurks. His sensorimotor connections to the world around him provides information on the rewards and penalties he needs to be well-adjusted – behave optimally, safely, curiously. Learning through interaction is the foundation of our existence. Equally, we can take a computational approach to this information interaction in the context of human and machine where now the roles are reversed. The machine is the human, and the human, the environment.

What would we need to understand in order to interact with an information system (machine) with our eyes or have the machine interact with us based on what it perceives in our eyes? Well of course, the machine would require a direct interface to an eye-tracking device which would provide a data stream. Consider gaze point as an example signal. What are the operational characteristics of this signal? Investigation of the speed and sensitivity of the signals is a fundamental objective for this interaction to make sense. Additionally, a human knows precisely when they wish to click, touch, or use their voice, to execute interactions. What can we say about the machine? How would the machine learn when to provide a context menu, retrieve a specific document, adjust the presentation, or filter the information? If such a system existed, how would we democratize it? As I will discuss later in great detail, Pupil Center Corneal Reflection (PCCR) eye tracking devices are extraordinarily expensive and thus research with them becomes self-limiting for building real-time adaptive systems as I have outlined above. 

Generally, the traditional methodology for information retrieval experiments has been to study gaze behavior and then report the findings in order to optimize interface layout or improve relevance feedback. If I were to ask \textit{where is the technology and how can I interact with it?} What we would find is that they are confined to aseptic laboratories. Sophisticated eye trackers utilize infrared illumination and pupil center corneal reflection methods to capture raw gaze coordinates and classify the ocular behavior as an all-in-one package. Local cues within highly visual displays of information are intended to be used to assess and navigate spatial relationships \cite{pirolli2001visual,pirolli2003effects}. Having functions that enable rapid, incremental, and reversible actions with continuous browsing and presentation of results are pillars of visual information seeking design \cite{ahlberg2003visual}. Moreover, \say{\textit{how does visualization amplify cognition?}} By grouping information together and using positioning intelligently between groups, reductions in search and working memory can be achieved and is the essence of \say{\textit{using vision to think}} \cite[see page 15-17]{card1999readings}. Thus, by studying ocular behavior of information retrieval processes, engineers can optimize their systems. This short review provides a historical background on ophthalmic neurophysiology, eye tracking technology, information retrieval experiments, and experimental considerations for those beginning work in this area.

\section{Ophthalmic Neurophysiology}
Millions of years of evolution through physical, chemical, genetic, molecular, biological, and environmental, pathways of increasing complexity naturally selected humans for something beautiful and fundamental to our senses and consciousness – visual perception. The knowledge gained since the first comprehensive anatomic descriptions of neural cell types that constitute the retina in the 19\textsuperscript{th} century followed by electron microscopy, microelectrode recording techniques, immunostaining, and pharmacology, in the 20\textsuperscript{th} century \cite{perlman2015organization} are immature in comparison to the forces of nature.

Now, here we are in the first-quarter of the 21\textsuperscript{st} century and human-machine interaction research scientists are asking the question \say{\textit{how can I leverage an understanding of vision and visual perception in my research and development process?}} As research scientists in the information field, we should bear this responsibility with conviction and depth to try and understand every possible angle of the phenomena we seek to observe and record. This section on \textit{Ophthalmic Neurophysiology} is an elementary introduction on how vision works and should be our prism through which we plan and execute all eye-tracking studies.

\begin{figure}[H]
		\centering
		\includegraphics[width=10cm]{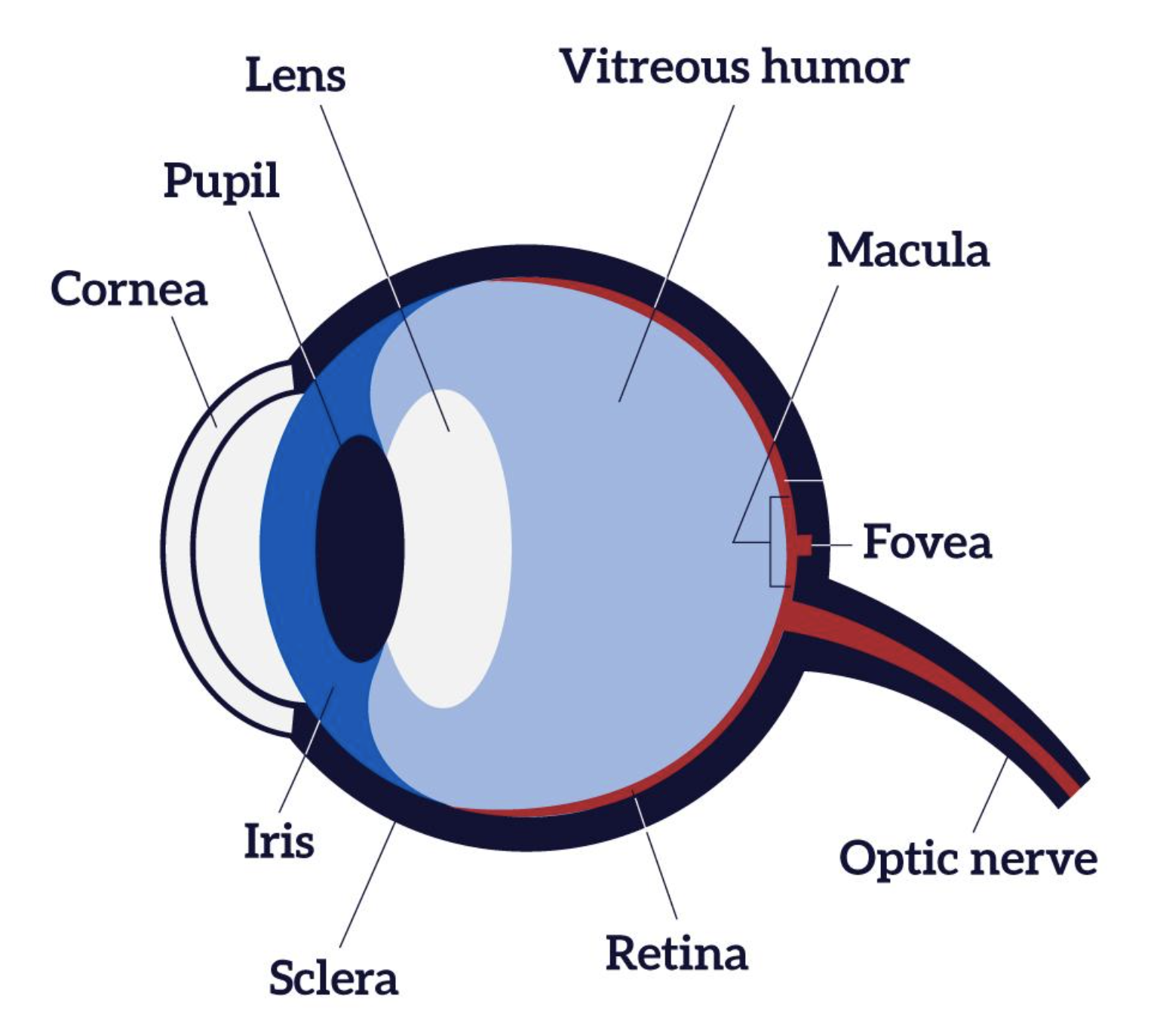}
		\caption{Basic eye anatomy. Credit: National Eye Institute.}
		\label{fig:fig1}
\end{figure}

Figure~\ref{fig:fig1} shows the basic anatomy of the eye. First, light passes through the cornea which due to its shape, can bend light to allow for focus. Some of this light enters through the pupil which has its diameter controlled by the iris. Bright light causes the iris to constrict the pupil which lets in less light. Low light causes the iris to widen the pupil diameter to let in more light. Then, light passes through the lens which coordinates with the cornea via muscles of the Ciliary body to properly focus the light on the light-sensitive layer of tissue called the retina. Photoreceptors then translate the light input into an electrical signal that travels via the optic nerve to the brain\footnote{\url{https://www.nei.nih.gov/learn-about-eye-health/healthy-vision/how-eyes-work}}.

\begin{figure}[H]
		\centering
		\includegraphics[width=8cm]{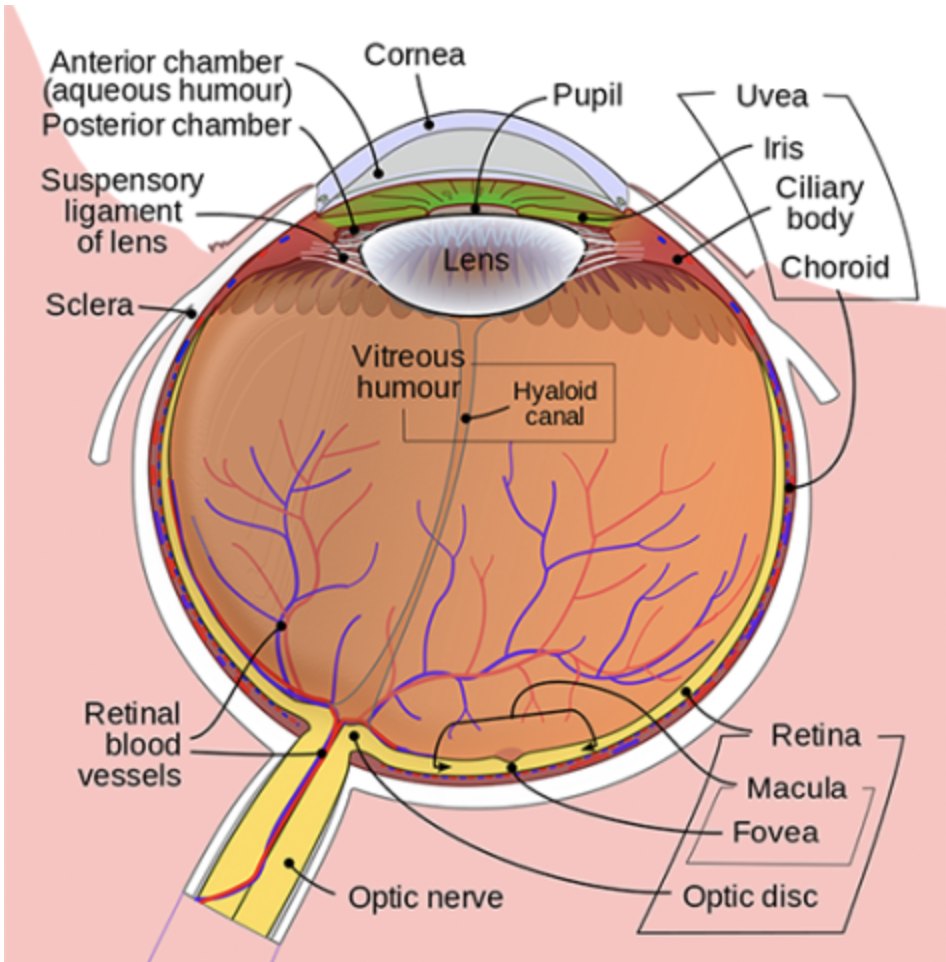}
		\caption{Complex eye anatomy. Credit: Kellogg Eye Center, Michigan Medicine.}
		\label{fig:fig2}
\end{figure}

Figure~\ref{fig:fig2} shows the slightly more complex anatomy of the eye as a cross-section. We will focus on the back of the eye (lower portion of the figure). The fovea is the center of the macula and provides sharp vision that is characteristic of attention on a particular stimulus in the world while leaving the peripheral vision somewhat blurred. You may notice the angle of the lens and fovea are slightly off-center. More on this later. The optic nerve is a collection of millions of nerve fibers that relay signals of visual messages that have been projected onto the retina from our environment to the brain.\footnote{\url{https://www.umkelloggeye.org/conditions-treatments/anatomy-eye}}

\begin{figure}[H]
		\centering
		\includegraphics[width=8cm]{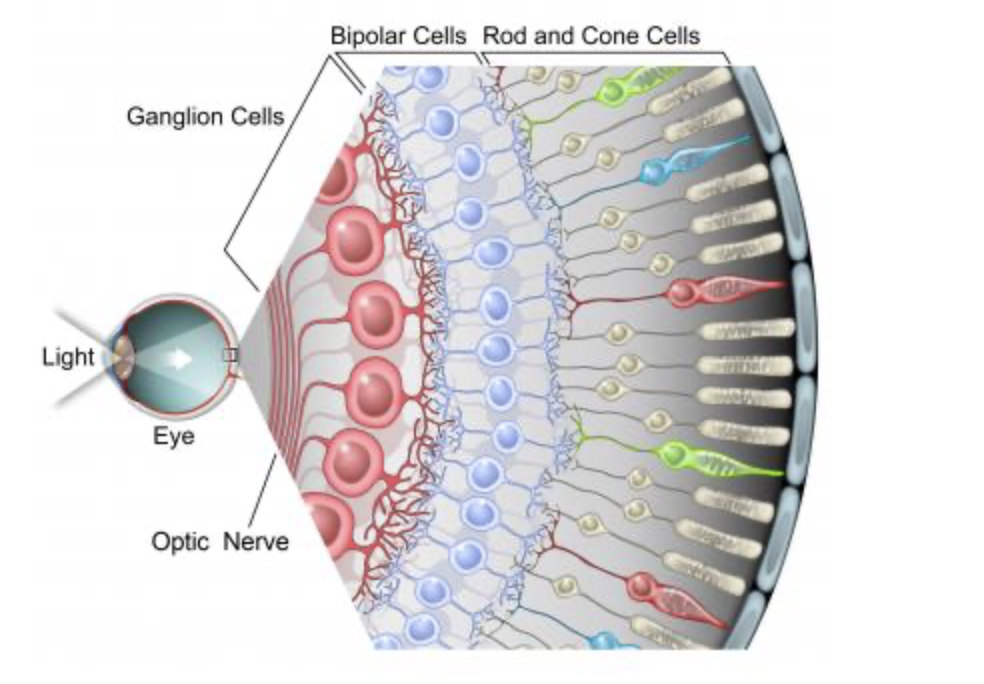}
		\caption{Neurons of the retina.}
		\label{fig:fig3}
\end{figure}

The electrical signals in transit to the brain first have to be spatially distributed across the five different neural cell types shown in figure~\ref{fig:fig3}. The photoreceptors (rods and cones) are the first order neurons in the visual pathway. These receptors synapse (connect and relay) with bipolar and horizontal cells which function primarily to establish brightness and color contrasts of the visual stimulus. The biploar cells then synapse with retinal ganglion and amacrine cells which intensify the contrast that supports vision for structure, shape, and is the precursor for movement detection. Finally the visual information that has been translated and properly organized into an electrical data structure is delivered to the brain via long projections of the retinal ganglion cells called axons.

\begin{figure}[H]
		\centering
		\includegraphics[width=8cm]{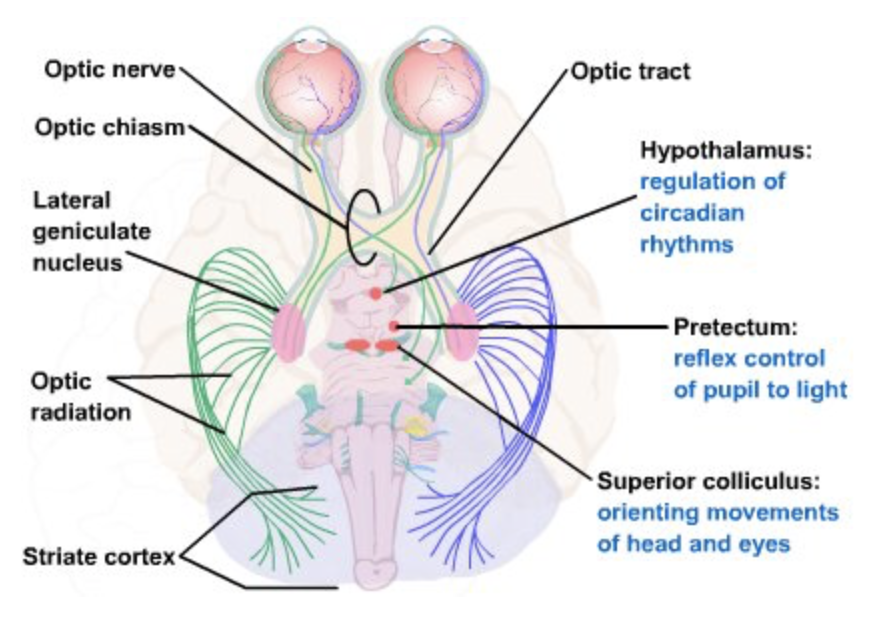}
		\caption{Simplified visual pathway; from retina to cortex. Credit: University of Texas McGovern Medical School.}
		\label{fig:fig4}
\end{figure}

Described thus far is, broadly, the visual pathway from external stimulus to retinal processing. Sensory information must reach the cerebral cortex (outer layer of the brain), to be perceived. We must now consider the visual pathway from retina to cortex as shown in the cross-section of figure~\ref{fig:fig4}. The optic nerve fibers intersect contralaterally at the optic chiasm. The axons in this \textit{optic tract} end with various nuclei (cell bodies). The thalamus is much like a hub containing nerve fiber projections in all directions that exchange information to the cerebral cortex (among many other regulatory functions).

\begin{figure}[H]
		\centering
		\includegraphics[width=10cm]{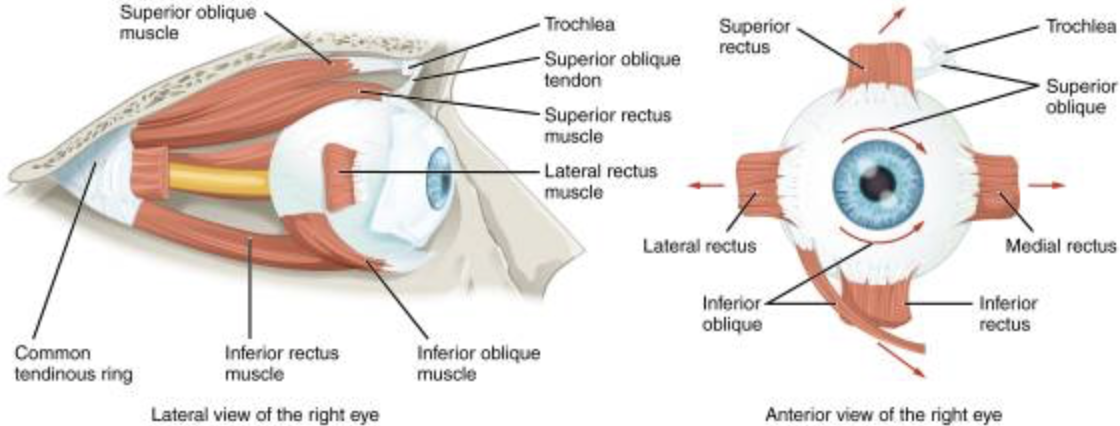}
		\caption{Lateral and anterior view of extraocular muscles \cite{ludwig2019anatomy}}
		\label{fig:fig5}
\end{figure}

\begin{figure}[H]
		\centering
		\includegraphics[width=8cm]{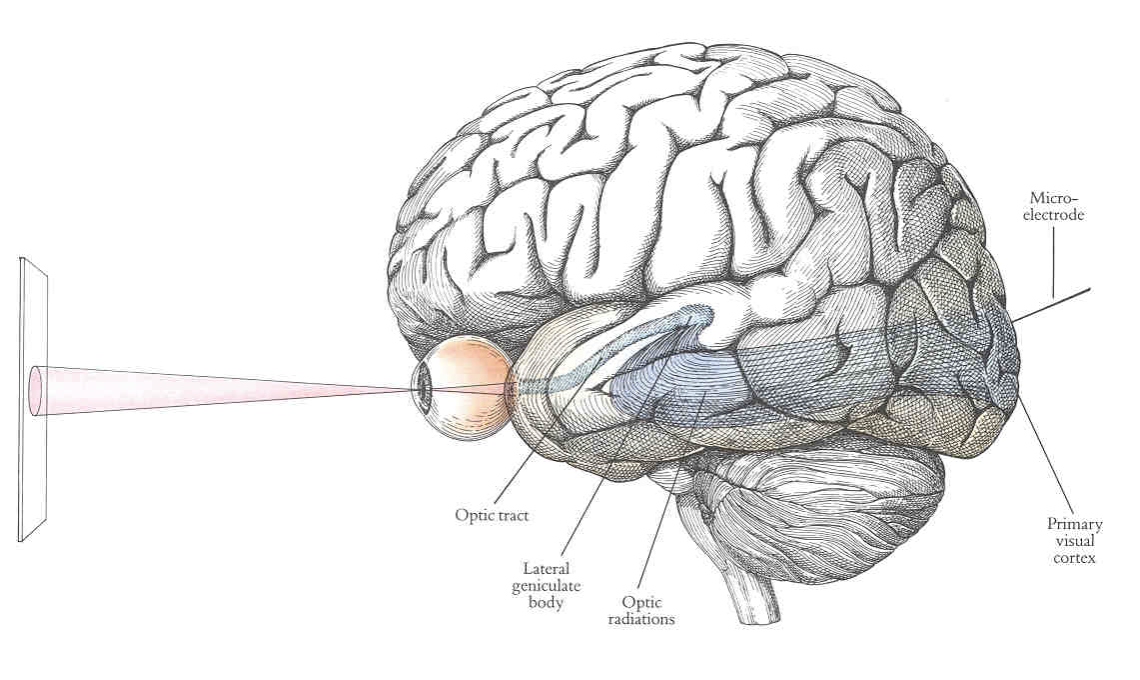}
		\caption{Lateral view of visual perception (Lateral Geniculate Nucleus). Credit: MIT Cortical Maps}
		\label{fig:fig6}
\end{figure}

Within the midbrain, involved in motor movements, there is the superior colliculus that plays an essential role in coordinating eye and head movements to visual stimuli (among other sensory inputs). For example, the extraocular muscles are shown in figure~\ref{fig:fig5}. Within the thalamus, the lateral geniculate nucleus coordinates visual perception\footnote{\url{http://www.mit.edu/~kardar/research/seminars/CorticalMaps/VisualSystem.html}} as shown in figure~\ref{fig:fig6}. Lastly, the pretectum controls the pupilary light reflex.\footnote{\url{https://nba.uth.tmc.edu/neuroscience/m/s3/chapter07.html}}

Based on the introductory ophthalmic neurophysiology reviewed in this section, human-machine interaction experimenters should consider (at a minimum) certain operating parameters:

\begin{itemize}
	\item Pupillary response to lighting conditions is sensitive. Control for this by maintaining stable lighting throughout an experiment, as one may not be able to defend that changes in pupil diameter are in-fact due to changes in focus/attention on the machine or changes in ambient lighting.
	\item Screen participants for no previous history of ophthalmic disease. If the visual system is impaired at any level, the neurophysiological responses are no longer a reliable dependent variable as an excitatory, lack of, or delayed, ophthalmic response may not accurately represent a neurophysiological transition state with respect to machine interaction.
	\item Many ophthalmic diseases are age-related. For the examination of human-machine interaction in the context of spatial/visual information, recruit study participants that are under the age of 40 to minimize the likelihood of confounding variables.\footnote{\url{https://www.cdc.gov/visionhealth/basics/ced/index.html}}\footnote{\url{https://www.nia.nih.gov/health/aging-and-your-eyes}}\footnote{\url{https://ghr.nlm.nih.gov/condition/age-related-macular-degeneration}}\footnote{\url{https://www.nei.nih.gov/learn-about-eye-health/resources-for-health-educators/vision-and-aging-resources}}\footnote{\url{https://www.aoa.org/patients-and-public/good-vision-throughout-life/adult-vision-19-to-40-years-of-age/adult-vision-41-to-60-years-of-age}}
\end{itemize}

After reviewing the itemized list above, some may reason that this preliminary screening criteria is too narrow due to the fact that neuroadaptive systems will soon emerge on the technological landscape and that aging populations are increasingly engaging with technology, therefore their neurophysiological responses should be studied in order to make technology inclusive, not exclusive. I happen to agree with this logic. However, as we will review later, many limitations in current measuring devices exist, and some are related to ophthalmic diseases or deficiencies.

\section{Eye-tracking Technology}
\begin{figure}[H]
		\centering
		\includegraphics[width=10cm]{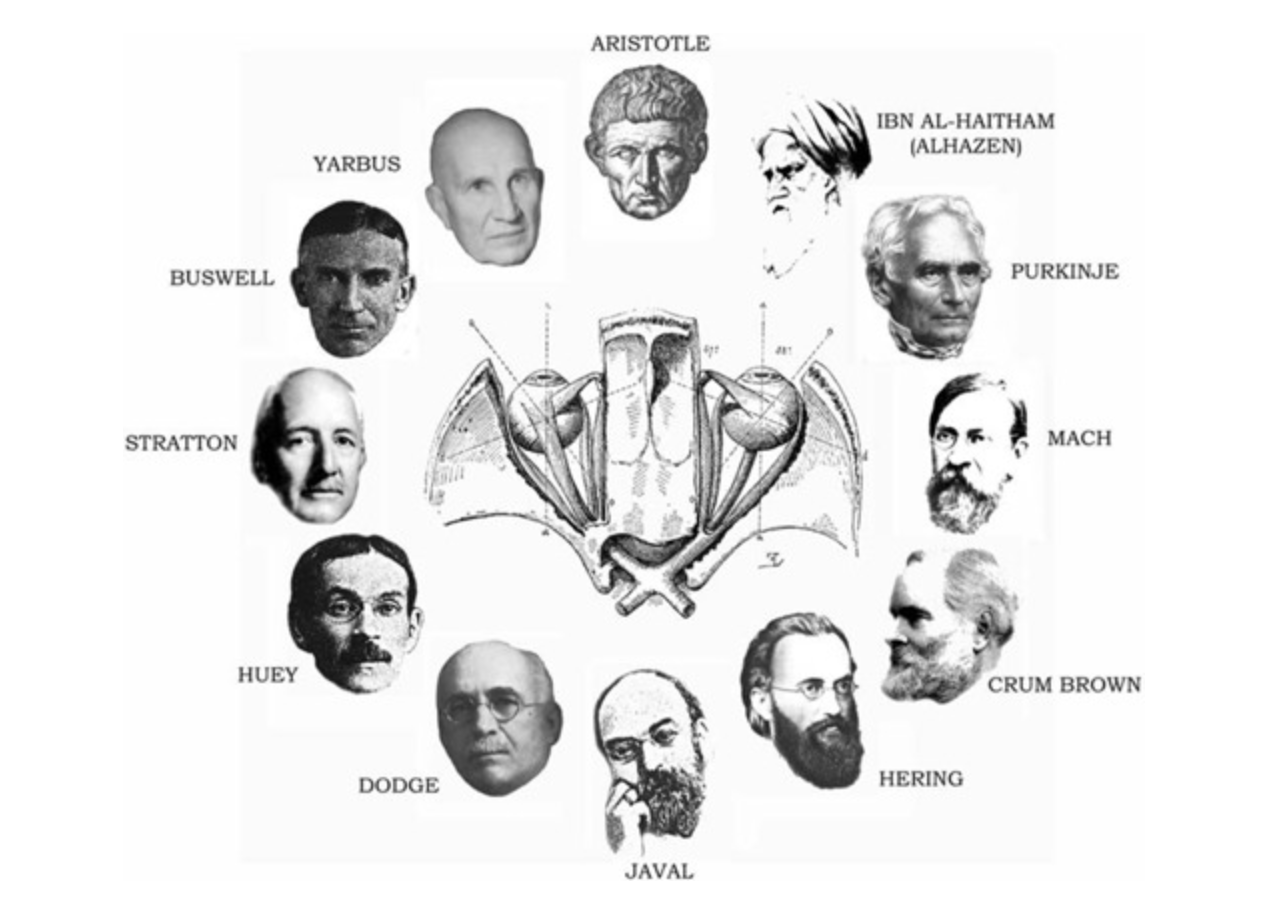}
		\caption{Pioneers of eye movement research.}
		\label{fig:fig7}
\end{figure}

In this section I will explain the history, theory, practice, and standardization of eye-tracking technology. The pioneers of eye-tracking date all the way back to Aristotle as can be seen in the clock-wise chronological arrangement in figure~\ref{fig:fig7}. Others contributed significantly to the knowledge of eye movement studies however portraits of them have either not been found or do not exist \cite{wade2010pioneers}.

In 1879, the discontinuities of eye movements were elucidated by the Swiss Ophthalmologist Edmond Landolt \cite{landolt1879manual}. Although his work did not use the terms \textit{fixations} and \textit{saccades} (rapid movement between fixations), it provided a framework for understanding the terms we use today. In the same year, the German physiologist Edwald Hering and French Ophthalmologist Louis \'Emile Javal, described the discontinuous eye movements during reading. Dr. Javal was an ophthalmic laboratory director at the University of Paris (Sorbonne), worked on optical devices, the neurophysiology of reading, and introduced the term \textit{saccades} which of Old French origin (8\textsuperscript{th} to 14\textsuperscript{th} century) was \textit{saquer} or \textit{to pull} and in modern French translates to \textit{violent pull}.

\vspace{5mm} 

\say{\textit{the eye makes several saccades during the passage over each line, about one for every 15–18 letters of text}} \cite{javal1878essai}. (French to English translation).

\vspace{5mm} 

About twenty years later, the psychologist Edmund Burke Huey appeared to be the first American to cite Javal's work describing that the consistent neurophysiological accommodation (referring to the lens of the eye) from having to read laterally across a page increases extraocular muscle fatigue and reduces reading speed \cite{huey1898preliminary}. Moreover, Dr. Huey described his motivations for building an experimental eye-tracking device:

\vspace{5mm} 

\say{\textit{the eye moved with along the line by little jerks and not with a continuous steady movement. I tried to record these jerks by direct observation, but finally decided that my simple reaction to sight stimuli was not quick enough to keep up... It seemed needful to have an accurate record of these movements; and it seemed impossible to get such record without a direct attachment of recording apparatus to the eye-ball. As I could find no account of this having been done, I arranged an apparatus for the purpose and have so far succeeded in taking 18 tracings of the eye's movements in reading.}}

\vspace{5mm} 

\begin{figure}[H]
		\centering
		\includegraphics[width=10cm]{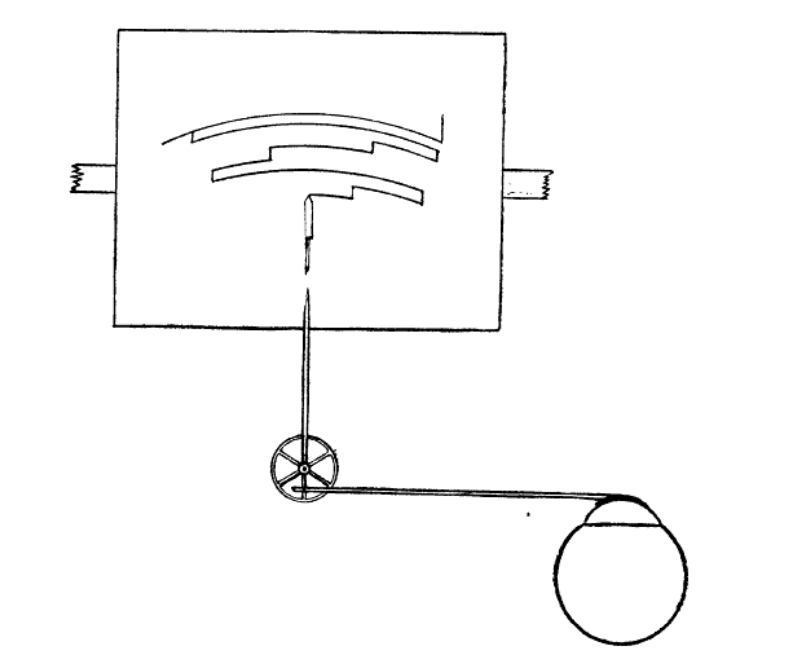}
		\caption{Huey eye-tracker.}
		\label{fig:fig8}
\end{figure}

A drawing of this apparatus is show in figure~\ref{fig:fig8}. Dr. Huey went on to write the famous book on \textit{Psychology and Pedagogy of Reading} in 1908 \cite{huey1908psychology}.

\begin{figure}[H]
	\centering
	\begin{minipage}[b]{0.4\textwidth}
		\includegraphics[width=\textwidth]{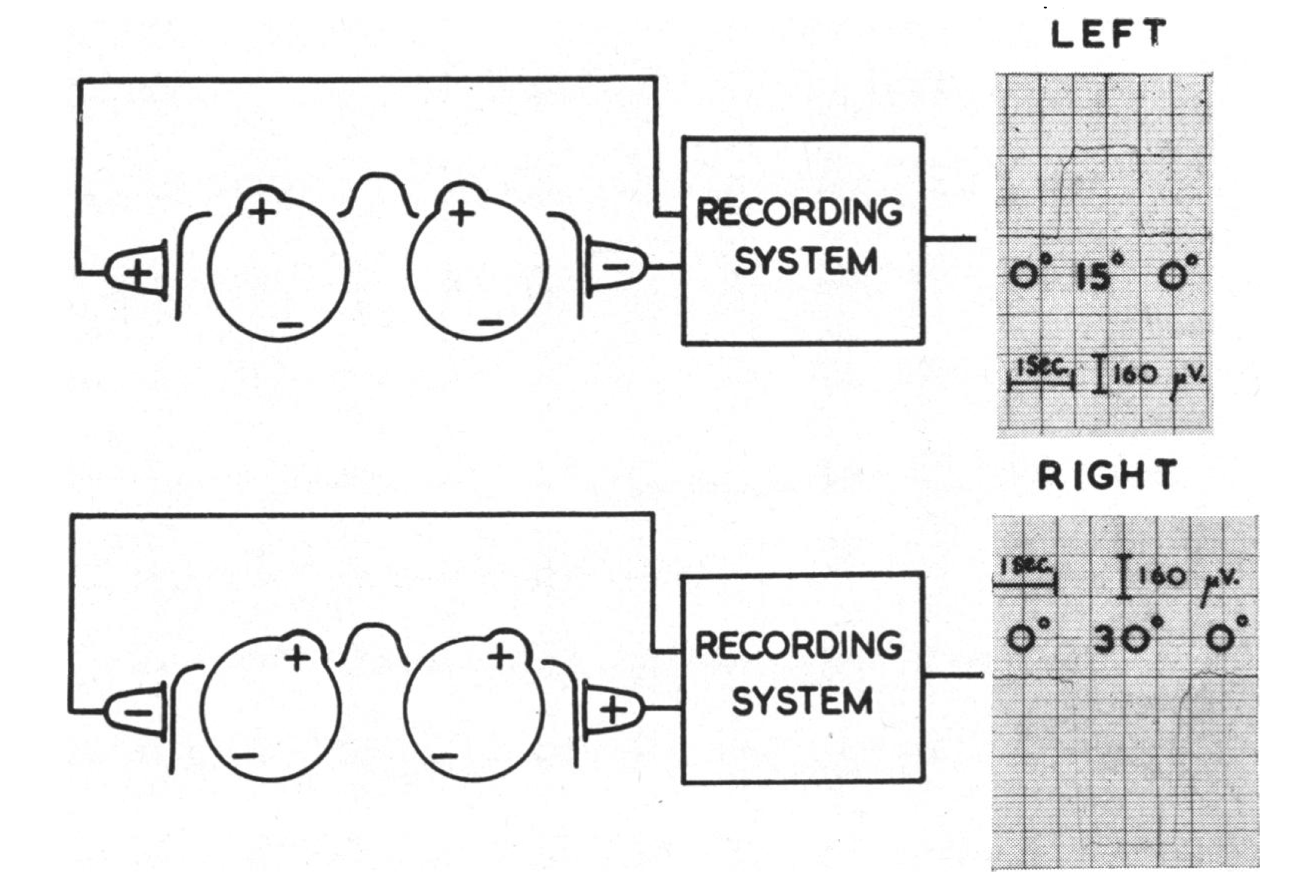}
		\caption{Paradigm.}
		\label{fig:fig9}
	\end{minipage}
	\hfill
	\begin{minipage}[b]{0.4\textwidth}
		\includegraphics[width=\textwidth]{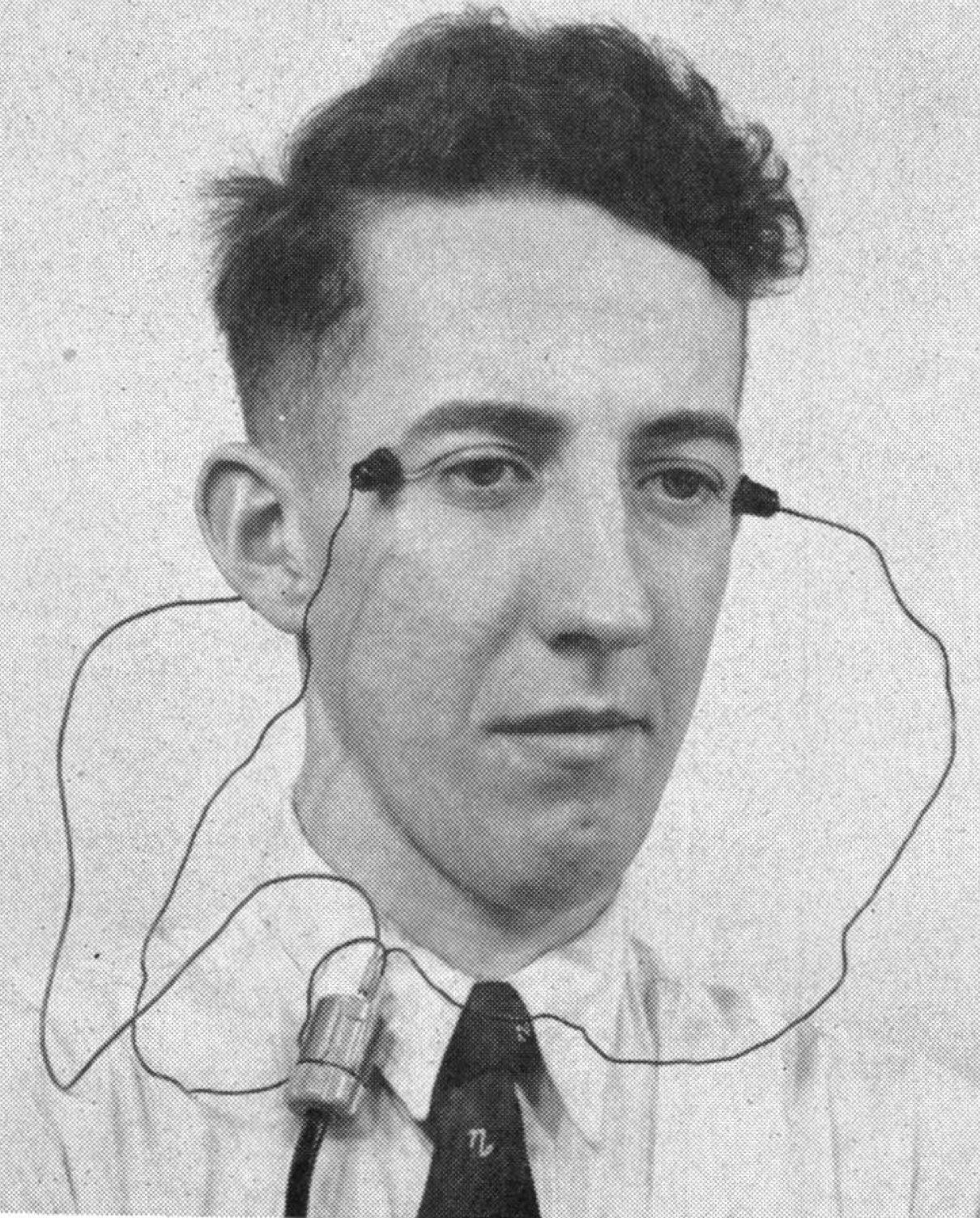}
		\caption{Electrodes.}
		\label{fig:fig10}
	\end{minipage}
		\hfill
	\begin{minipage}[b]{0.4\textwidth}
		\includegraphics[width=\textwidth]{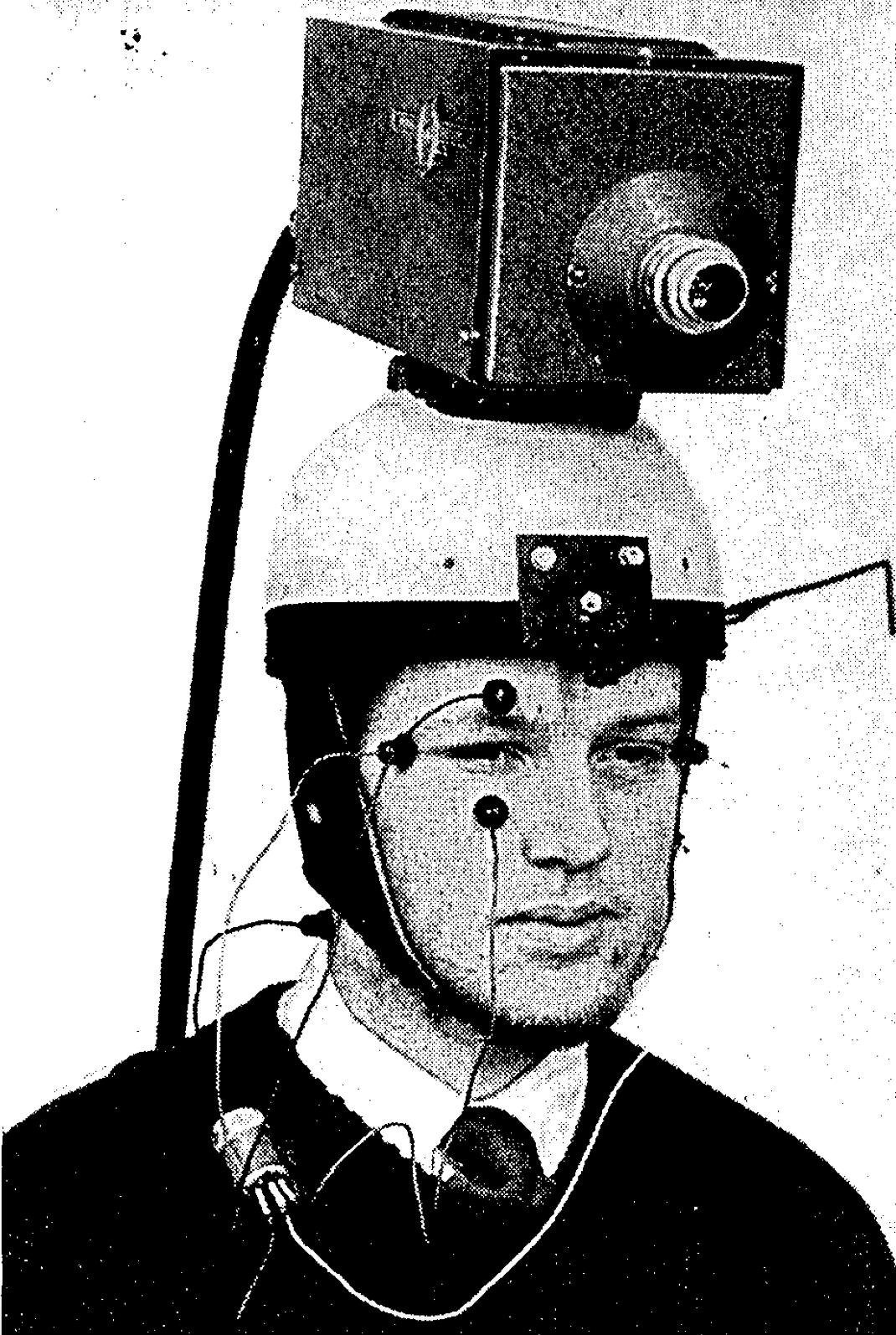}
		\caption{Visual Field.}
		\label{fig:fig11}
	\end{minipage}
\end{figure}

For an excellent historical overview of eye-tracking developments in the study of fixations, saccades, and reading, in the 19\textsuperscript{th} and 20\textsuperscript{th} centuries please see sections 6 and 6.1 in \cite{wade2010pioneers}. Additionally, and of particular interest, is the work in the 1960's of the British engineering psychologist, B. Shackel, who worked on the inter-relation of man and machine and the optimum design of such equipment for human use. Specifically his early work in measures and viewpoint recording of electro-oculography (electrical potential during eye rotation) for the British Royal Navy on human-guided weapon systems \cite{shackel1960note,shackel1960pilot} (see \cref{fig:fig9,fig:fig10,fig:fig11}).

\begin{figure}[H]
		\centering
		\includegraphics[width=5cm]{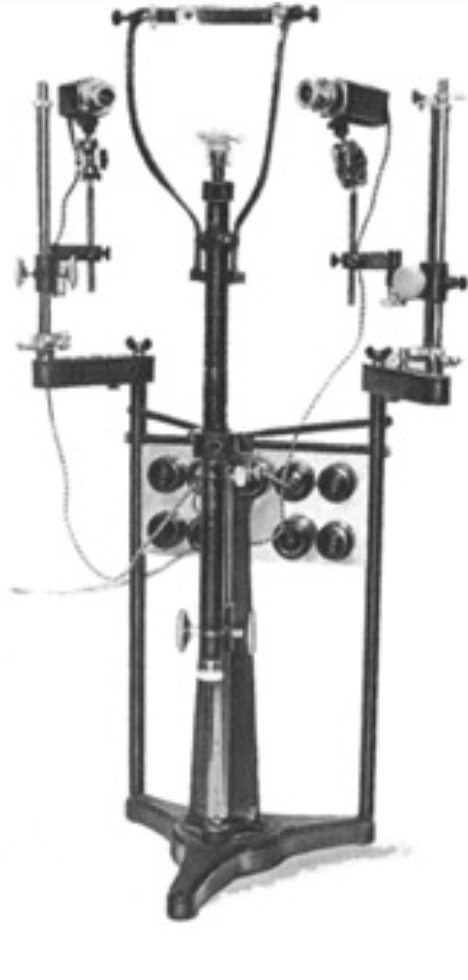}
		\caption{Yarbus suction cups and recording device.}
		\label{fig:fig12}
\end{figure}

The Russian psychologist Alfred L. Yarbus studied the relation between fixations and interest during image studies that used a novel device developed in his laboratory (figure~\ref{fig:fig12}). Please see Chapter IV in \cite{yarbuseye} for a thorough review of his experiments.

Merchant et al. were American engineers who worked for the aerospace systems arm of Honeywell International Inc. in the 1970's and developed a remote video oculometer for the United States Air Force that was prototyped to use eye movement as a control device for targeting software within aircraft weapon systems \cite{merchant1974remote}. The fundamental breakthrough of their research was digital image capture, automatic image processing, gaze point detection, and control device mapping, as a packaged hardware/software solution all in \textit{real-time}. Moreover, the basic concept of using a light source to illuminate the eye, create reflections, and capture an image of the eye showing such reflections in order to record real-time neurophysiologic phenomena, was called Pupil Center Corneal Reflection (PCCR) and is still the fundamental technology for state-of-the-art eye-trackers today, in the year 2020.\footnote{\url{https://www.tobiipro.com/learn-and-support/learn/eye-tracking-essentials/how-do-tobii-eye-trackers-work/}} Although, the design as it related to form factor, dark room requirements, and restriction of head movement, were sub-optimal for \say{use in the wild}. Unfortunately, as history has shown us, when mission critical United States military funded research projects fail on deliverables, the research community follows in its abandonment of theory and practice and thus many years passed before innovations in eye-tracking emerged once again. However, metrics of performance was the overarching contribution by the early pioneers and are not limited to:

\begin{itemize}
		\item Pupil and iris detection.
		\item Gaze prediction.
		\item Freedom of head movement.
		\item Adjustments for human anatomical eye variability.
		\item Adjustment for uncorrected and corrected human vision.
		\item Environmental lighting.
		\item Ease of calibration.
		\item Safety (apparatus, exposure, etc.).
		\item Speed of capture, processing, prediction.
		\item Form factor and cost.
\end{itemize}

\begin{figure}[H]
		\centering
		\includegraphics[width=7cm]{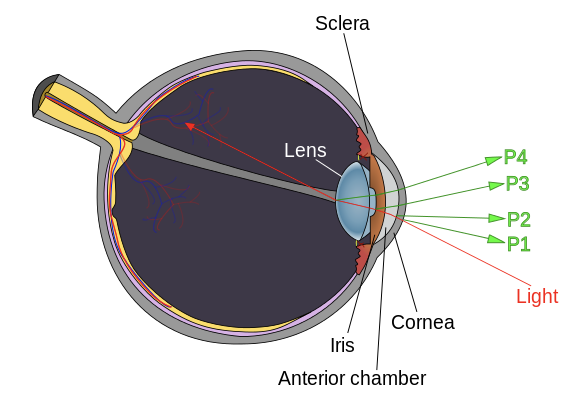}
		\caption{Light and the Purkinje images/reflections.}
		\label{fig:fig13}
\end{figure}

Let's discuss the Pupil Center Corneal Reflection (PCCR) method in more detail. Near-infared illumination creates reflection patterns on the cornea and lens called Purkinje images \cite{gellrich2016purkinje} (see figure \ref{fig:fig13}) which can be captured by image sensors and the resulting vectors can be calculated in real-time that describe eye-gaze and direction. This information can be used to analyze the behavior and consciousness of a subject \cite{elvesjo2009method}.

\begin{figure}[H]
		\centering
		\includegraphics[width=12cm]{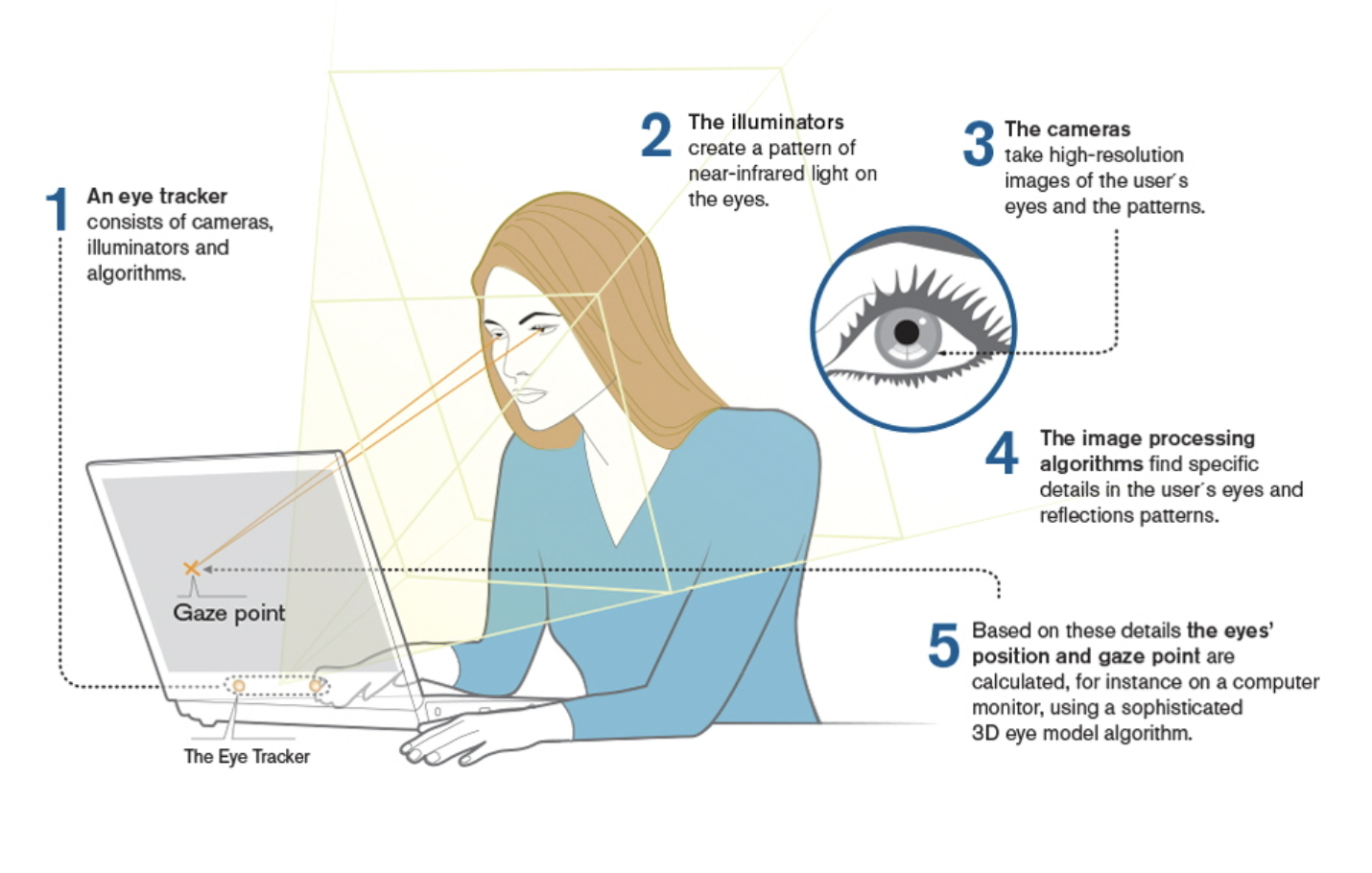}
		\caption{Tobii screen-based eye tracker architecture. Credit: Tobii Technology LLC.}
		\label{fig:fig14}
\end{figure}

The essential architecture of an eye-tracking device is made up of illuminators, cameras, processing unit for image detection, and a 3D eye model with fixation, saccade, and pupil size variation mapping algorithms. Figure~\ref{fig:fig14} demonstrates the paradigm for a Tobii Technology screen based eye tracker.\footnote{\url{https://www.tobiipro.com/learn-and-support/learn/eye-tracking-essentials/how-do-tobii-eye-trackers-work/}} Pupil response as an indicator of a neurophysiological state requires the establishment of a baseline pupil diameter. Additionally, pupil variations over time is the important marker versus pupil size, and when used in conjunction with ocular-motor changes over time (fixations and saccades), can provide a rich representation of human-machine interaction.\footnote{\url{https://www.tobiipro.com/learn-and-support/learn/eye-tracking-essentials/is-pupil-size-calculations-possible-with-tobii-eye-trackers/}}

\begin{figure}[H]
		\centering
		\includegraphics[width=9cm]{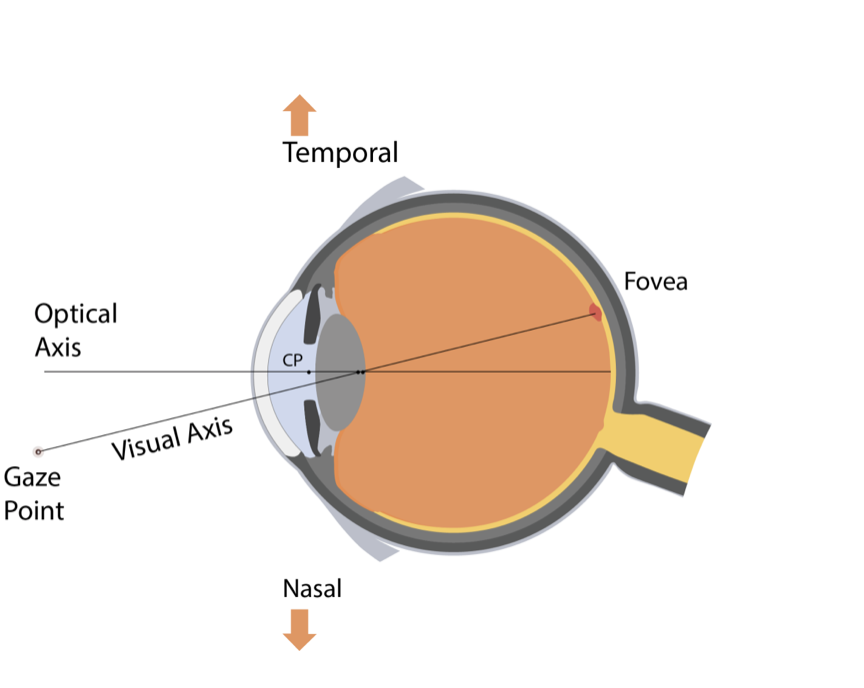}
		\caption{Geometrical variation calibration. Credit: Tobii Technology LLC.}
		\label{fig:fig15}
\end{figure}

Lastly, geometric characteristics of a subject's eyes must be estimated to reliably measure eye-gaze point calculations (see figure~\ref{fig:fig15}). Therefore, a calibration procedure involves bright/dark pupil adjustments for lighting conditions, light refraction/reflection properties of the cornea, lens, and fovea, and an anatomical 3D eye model to estimate foveal location responsible for the visual field (focus, full color).\footnote{\url{https://www.tobiipro.com/learn-and-support/learn/eye-tracking-essentials/what-happens-during-the-eye-tracker-calibration/}}

\section{Eye-tracking in Search and Retrieval}

In 2003, the first study on eye-tracking and information retrieval (IR) from search engines was conducted \cite{SalojarviJarkko}. The authors of the study wanted to understand if it was possible to infer relevance from eye movement signals. In 2004, Granka et al. \cite{10.1145/1008992.1009079} investigated how users interact with search engine result pages (SERPs) in order to improve interface design processes and implicit feedback of the engine while Kl\"{o}ckner et al. \cite{10.1145/985921.986115} asked the more basic question of search list order and eye movement behavior to understand depth-first or breadth-first retrieval strategies. In 2005, similar to the previous study, Aula et al. \cite{10.1007/11555261_104} wanted to classify search result evaluation \textit{style} in addition to depth-first or breadth-first strategies. The research revealed that users can be categorized as \textit{economic} or \textit{exhaustive} in that the eye-gaze of experienced users is fast and decisions are made with less information (economic).

In 2007, Cutrell and Guan \cite{10.1145/1240624.1240690} approached eye-tracking methodology in information retrieval a bit differently. They argued that search engine interfaces were remarkably similar across domains and that the metadata (e.g. title, snippets, URL, date, author, etc.) describing each result, although very simple in design, it was not obvious how users utilized this descriptive data. Essentially, what were they looking at when they made their decisions about a particular result item and what separates an expert searcher from a novice? Later, this research was complemented by the 2009 work of Ishita et al. \cite{10.1145/1555400.1555485} where it was observed that titles of search results explained much of the eye-tracking data, while Capra et al. used eye-tracking to examine how exploratory search within online public access catalogs (OPAC) was conducted during utilization of facets for filtering and refining a non-transactional search strategy \cite{capra2009faceted}.

In 2010, Balatsoukas and Ruthven \cite{doi:10.1002/meet.14504701145} argued that \say{there are no studies exploring the relationship between relevance criteria use and human eye movements (e.g. number of fixations, fixation length, and scan-paths)}. I believe their was some truth to this statement, as the only research close to their work was that of inferring relevance, at the macro-level, from eye-tracking \cite{SalojarviJarkko}. Their work uncovered that topicality explained much of the fixation data. Dinet et al. \cite{10.1145/1941007.1941022} studied visual strategies of young people from grades 5 to 11 on how they explored the search engine results page and how these strategies were affected by typographical cuing such as font alterations while Dumais et al. \cite{10.1145/1840784.1840812} examined individual differences in gaze behavior for all elements on the results page (e.g. results, ads, related searches).

In 2012, Balatsoukas and Ruthven extended their previous work on the relationship between relevance criteria and eye-movements to include cognitive and behavioral approaches with grades of relevance (e.g. relevant, partial, not) and the relationship to length of eye-fixations \cite{10.1002/asi.22707} while Marcos et al. \cite{10.5555/2377916.2377949} studied patterns of successful vs. unsuccessful information seeking behaviors; specifically, how, why, and when, users behave differently with respect to query formulation, result page activity, and query re-formulation. In 2013, Maqbali et al. studied eye-tracking behavior with respect to textual and visual search interfaces as well as the issue of data quality (e.g. noise reduction, device calibration) at a time when the existing software\footnote{\url{https://www.tobiipro.com/learn-and-support/learn/tobii-pro-studio/}} did not support such features \cite{10.1145/2537734.2537747}.

In 2014, Gossen et al. studied the differences in perception of search results and interface elements between late-elementary school children and adults with the goal of developing methodologies to build search engines for engaging and educating young children based on previous evidence that search behavior varies widely between children and adults \cite{10.1145/2556288.2557031}. Gwizdka examined the relationship between the degree of relevance assigned to a retrieval result by a user, the cognitive effort committed to reading the documented result, and inferring the relationship with eye-movement patterns \cite{10.1145/2578153.2578198} while Hofmann et al. examined interaction and eye-movement behavior of users with query auto completion rankings (also referred to as query suggestions or dynamics queries)  \cite{10.1145/2661829.2661922}.

In 2015, Eickhoff et al. argued that query suggestion approaches were \say{attention oblivious} in that without mapping mouse cursor movement at the term-level of search engine result pages, eye-tracking signals, and query reformulations, efforts of user modeling were limited in their value, based solely on previous, popular, or related searches, and not entirely obvious that such suggestions were relevant for users with non-transactional information needs \cite{10.1145/2766462.2767703}. Gwizdka and Zhang examined relevance from the perspective of pupillary responses (pupil dilation) of users during visits and re-visits of web pages and hypothesized differences in pupil dilation would reflect relevance, as such physiologic responses may indicate level of interest and can be used as a proxy for relevance \cite{10.1145/2766462.2767795} while Liu et al. studied eye-movements and cursor activity in the context of vertical search sessions and how vertical type and position, within the results ranking, affected neurophysiologic and behavioral measures \cite{10.1145/2766462.2767714}.

In 2016, Bilal and Gwizdka re-examined the reading behavior of children, grades 6 and 8, by asking what eye-fixation patterns can be observed by manipulating task type for Google SERPs with the vision of developing child-centric models of readability for improved access and comprehension \cite{10.5555/3017447.3017536} while Mostafa and Gwizdka extended the notion of controlled experimentation to go beyond that of implemented systems that have variations for experimentation such as \textit{control} and \textit{experimental} conditions, to that of neurophysiological baselines as an important step in the experimental design process and analogize the biological and clinical \textit{bio-marker} to that of:

\vspace{5mm} 

\say{\textit{establish behavioral correlates or markers that indicate normal or abnormal psycho-physiological conditions}}

\vspace{5mm} 

...in order to contextualize experimental responses  \cite{Mostafa2016DeepeningTR}. Prior to their position, experimental concerns were focused on data quality (e.g. noise reduction) and device calibration, not human response calibration.

In 2017, Gwizdka et al. revisited previous work on inferring relevance judgements for news stories albeit with a higher resolution eye-tracking device and the addition of more complex neurophysiological approaches such as electroencephalography (EEG) to identify relevance judgement correlates between eye-movement patterns and electrical activity in the brain \cite{10.5555/3204593.3204595} while Low et al. applied eye-tracking, pupillometry, and EEG to model user search behavior within a multimedia environment (e.g. an image library) in order to operationalize the development of an assistive technology that can guide a user throughout the search process based on their predicted attention, and latent intention \cite{10.1145/3020165.3022131}.

In 2019, the first neuroadaptive implicit relevance feedback information retrieval system was built and evaluated by Jacucci et al. \cite{doi:10.1002/asi.24161}. The authors demonstrated how to model search intent with eye and brain-computer interfaces for improved relevance predictions while Wu et al. examined eye-gaze in combination with electrodermal activity (EDA), which measures neurally mediated effects on sweat gland permeability, while users examined search engine result pages to predict subjective search satisfaction \cite{doi:10.1002/asi.24240}. In 2020, Bhattacharya et al. re-examined relevance prediction for neuroadaptive IR systems with respect to scanpath image classification and reported up to 80\% accuracy in their model \cite{bhattacharya2020relevance}.

\section{Eye-tracking in Aware and Adaptive User Interfaces}

In this section, I will review only those works that satisfy the criteria of a system (machine) that utilizes implicit signals from an eye-tracker to carry out functions and interact or collaborate with a human.

iDict was an eye-aware application that monitored gaze path (saccades) while users read text in foreign languages. When difficulties were observed by analyzing the discontinuous eye movements, the machine would assist with the translation \cite{10.1145/355017.355019}. Later, an affordable \say{Gaze Contingent Display} was developed for the first time that was operating system and hardware integration agnostic. Such a display was capable of rendering images via the gaze point and thus had applications in gaze contingent image analysis and multi-modal displays that provide \say{focus+context} as can be found with volumetric medical imaging \cite{10.1145/968363.968366}.

Children with autism spectrum disorder have difficulties with social attention. Particularly, they do not focus on the eyes or faces of those communicating with them. It is thought that forms of training may offer benefit. An amusement ride machine was engineered and outfitted with various sensors and an eye-tracker. The ride was an experiment that would elicit various responses from the child and require visual engagement of a screen that would then reward with auditory and vestibular experiences, and thus functioned as a gaze contingent environment for socially training the child on the issue of attention \cite{10.1145/968363.968367}.

Fluid and pleasant human communication requires visual and auditory cues that are respected by two or more people. For example, as I am speaking to someone and engaged in eye contact, perhaps I will look away for a moment or fade my tone of voice and pause. These are social cues that are then acted upon by another person where they then engage me with their thought. This level of appropriateness is not embedded in devices. Although the concept of \say{Attentive User Interfaces} that utilize eye-tracking to become more conscious about when to interrupt a human or group of humans has been studied \cite{10.1145/968363.968384}.

Utilizing our visual system as a point and selection device for machine interactions instead of a computer mouse or touch screen would seem like a natural progression in the evolution of interaction. There are two avenues of engineering along this thread. The first simply requires a machine to interact with, an accurate eye tracking device, and thresholds for gaze fixation in order to select items presented by the machine. The second requires that we study the behaviors of interaction (eye, peripheral components) and their correlates in order to build a model of what the typical human eye does precisely before and after selections are made. 

With this information we may then be able to have semi-conscious machines that understand when we would like to select something or navigate through an environment. A machine of the first kind was in-fact built and experimented on for image search and retrieval \cite{10.1145/1117309.1117324}, whereby a threshold of 80 millisecond gaze fixation was used as the selection device. The experiment asked that users identify the target image within a library of images that were presented in groups. All similarity calculations were stored as metadata prior to the experiment. The user would have to iteratively gaze at related images for at least 80 milliseconds for the group of images to filter and narrow with a change of results. The results indicated that the speed of gaze contingent image search was faster than an automated random selection algorithm. However, the gaze contingent display was not experimented against a traditional interaction like the computer mouse. 

Later, a similar system was built and an experiment was conducted using Google image search \cite{10.1145/1743666.1743684}. The authors in \cite{10.1145/2858036.2858137} also presented a similar gaze threshold (100 ms) based system called \textit{GazeNoter}. The gaze-adaptive note-taking system was built and tested for online PowerPoint slide presentations. Essentially, by analyzing a user's gaze, video playblack speed would adjust and recommend notes for particular areas of interest (e.g. bullet points, images, etc.). The prototype was framed around the idea that video lectures require the user to obtrusively pause the video, lose focus, write a note, then continue. In-fact, the experiments reported show that users generated more efficient notes and preferred the gaze adaptive system in comparison to a baseline system that had no adaptive features.

In \cite{10.1145/2857491.2888590}, the authors note that implementation of eye-tracking in humanoid robots has been done before. However, no experiment had been conducted on the benefits for human-robot collaboration. The eye-tracker was built into the humanoid robot \say{iCub}\footnote{\url{http://www.icub.org}} as opposed to being an externally visible interface. This engineering design enabled a single blind experiment where the subjects had no knowledge of any infared cameras illuminating the cornea and pupil or the involvement of eye-tracking in the experiment. The robot and human sat across each other at a table. The humans were not asked to interact with the robot in any particular way (voice, pointing, gaze, etc.) but were asked to communicate with the robot in order to receive a specific order of toy blocks to build a structure. The robot was specifically programmed in this experiment to only react to eye gaze which it did successfully in under 30 seconds across subjects.

Cartographers encode geographic information on small scale maps that represent all the topological features of our planet. This information is decoded with legends that enable the map user to understand what and where they are looking at. Digital maps have become adaptive to user click behavior and therefore the legends reflect the real-time interaction. Google Earth\footnote{\url{https://www.google.com/earth/}} is an excellent example of this. New evidence indicates that gaze-based adaptive legends are just as useful and perhaps more useful than traditional legends \cite{10.1145/3204493.3204544}. This experiment included two versions of a digital map (e.g. static legend, gaze-based adaptive legend). Although participants in the study performed similarly for time-on-task, they preferred the adaptive legend, indicating its perceived usefulness.

\section{Additional Experimental Considerations}
\subsection{Technology}
The standardization of eye-tracking technology is not without limitation. A number of advancements in the fundamental technology of PCCR based eye-trackers are still required. For example, the image processing algorithms have difficulty on a number of scenarios involving the pupil center corneal reflection method:

\begin{itemize}
		\item Reflections from eye-glasses and contact lenses worn by the subject can cause image processing artifacts.
		\item Eye-lashes that occlude the perimeter of the pupil cause problems for time-series pupil diameter calculations.
		\item Large pupils reflect more light than small pupils. The wide dynamic range in reflection can be an issue for image processors.
		\item The eye blink reflex has a complex neural circuit involving the oculomotor nerve (cranial nerve III), trigeminial nerve (cranial nerve V), and the facial nerve (cranial nerve VII). \footnote{\url{https://nba.uth.tmc.edu/neuroscience/m/s3/chapter07.html}}\footnote{\url{https://www.ncbi.nlm.nih.gov/books/NBK534247/}} When a pathology in this reflex is present the subject does not blink during an experimental task therefore dry and congealed corneas is the result, which makes corneal reflection difficult for the image processor.
		\item High-speed photography by the image capture modality is required as saccadic eye movements have high velocity, and head movements may at times be also high in velocity causing blurred images of the corneal reflection.
		\item Squinting causes pupil center and corneal reflection distortion during image processing.
		\item The trade-off between PCCR accuracy and freedom of head movement may be overcome by robotic cameras that \say{\textit{eye follow}} although this is not available in most affordable eye-trackers.
\end{itemize}

\begin{figure}[H]
		\centering
		\includegraphics[width=10cm]{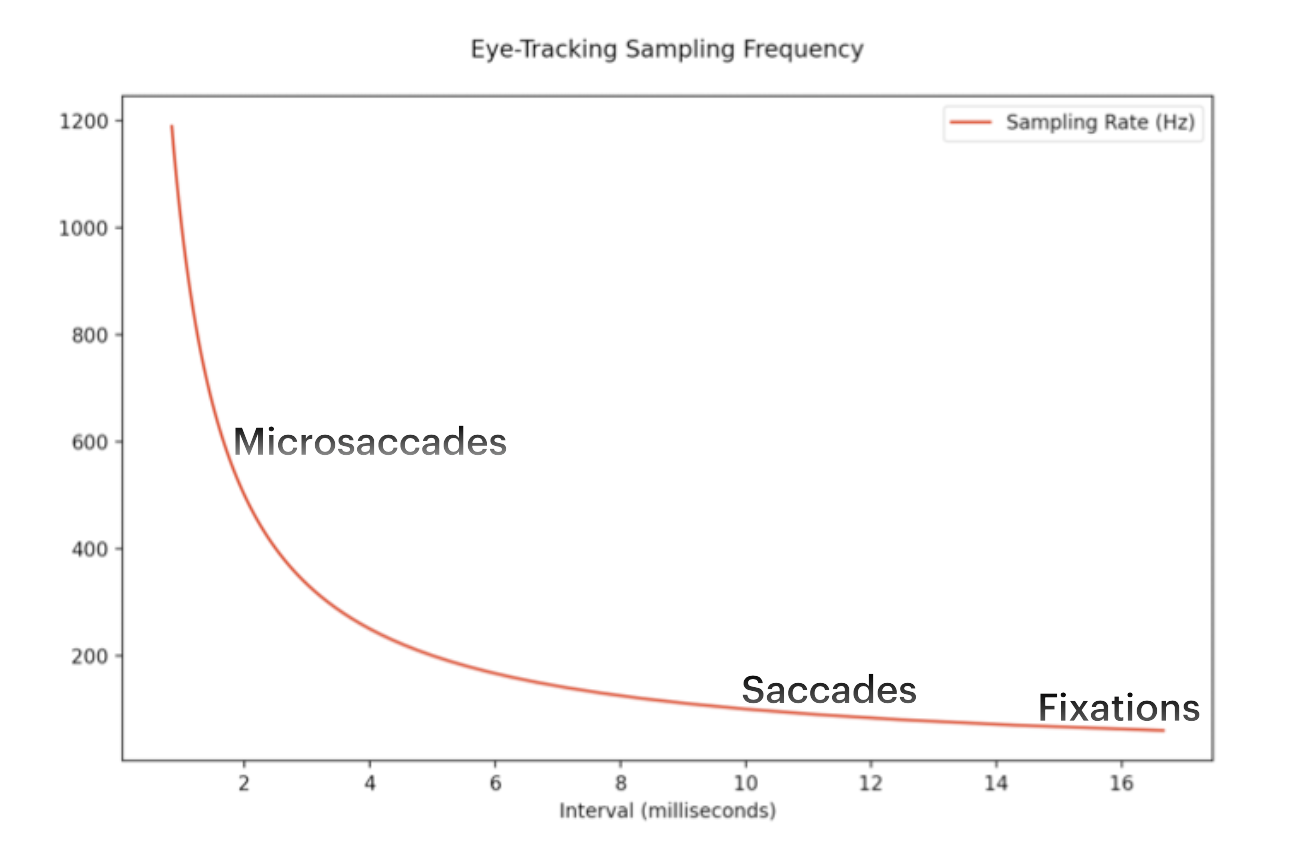}
		\caption{Sampling frequencies.}
		\label{fig:fig16}
\end{figure}

Additionally, sampling frequencies should be thoughtfully understood in order to design an experiment that potentially answers a question or set of questions (see figure~\ref{fig:fig16}). Essentially, at the highest frequency (1200 Hz), 1200 data points for each second of eye movement are recorded and each eye movement will be recorded approximately every 0.83 milliseconds (sub-millisecond). While at the lowest end of the frequency spectrum (60 Hz), 60 data points for each second of eye movement are recorded and each eye movement will be recorded approximately every 16.67 milliseconds. These sampling frequencies are important to understand because certain eye phenomena can only be observed at certain frequencies. For example, low-noise saccades are observed at frequencies greater than 120Hz which are sampled every 8.33 milliseconds while low-noise microsaccades are observed at frequencies greater than 600Hz which are sampled every 1.67 milliseconds.\footnote{\url{https://www.tobiipro.com/learn-and-support/learn/eye-tracking-essentials/eye-tracker-sampling-frequency/}}

Higher sampling frequencies will provide higher sample sizes and levels of certainty over the same unit of time. In terms of stratifying a data stream accurately and building user models for adaptive feedback within a system, high sampling frequency is a pre-requisite and provides more granularity for fixations, fixation duration, pupil dilation, saccades, saccade velocity, microsaccades, and spontaneous blink rate.

\subsection{Psychophysiology}

With this data, we can begin to ask questions related to moment-by-moment actions and their relationship to neurophysiology. For example, it is not possible to move your eyes (voluntarily or involuntarily) without making a corresponding shift in focus/attention and disruption to working memory. This is especially true in spatial environments \cite{shepherd1986relationship, theeuwes2009interactions}. Perhaps, by modeling a user's typical pattern of eye movement over time, a system can adapt and learn when to politely re-focus the user and/or more accurately model the eye-as-an-input.

Moreover, the eyes generally fixate on objects of thought. Although this may not always be the case in scenarios where we are looking at nothing but retrieving representations in the background \cite{ferreira2008taking}. Think of a moment where you gestured with your hand at a particular area of a room or place that someone you spoke to earlier was in. Therefore, in the context of a human-machine interaction, how would the machine learn to understand the difference in order to execute system commands, navigate menus, or remain observant for the next cue? For information systems, at least, this is the argument for supplementary data collection from peripheral components which allow for investigation and potential discovery of correlates that the machine can be trained on to understand the difference. However, an accepted theory of visual perception is that it is the result of both feedforward and feedback connections, where the initial feedforward stimulation generates interpretation(s) that are fed-backward for further processing and confirmation known as a reentrant loop. Experiments have demonstrated varying cycle times for reentrant loops when subjects are presented with information in advance (a specific task) for sequential image processing and detecting a target. Detection performance increased as the duration of an image being presented increased from 13-80 milliseconds \cite{potter2014detecting}. 

Another limitation with this interaction is the manipulation device (computer mouse) as the literature has suggested that average mouse pointing time for web search appears to range from 600-1000 milliseconds \cite{murata2009basic} while pupil dilation can have latencies of only 200 milliseconds. This suggests that visual perception during information seeking tasks is significantly faster than the ability to act on it with our motor movements and thus it is likely that the eye-as-an-input device is more efficient and therefore a significant delay between the moment a user decides upon a selection item and when the selection item is actuated, appears to exist. On this particular issue, experimental protocols should outlines a specific manner in which to understand or operationalize this gap.

Even when a user is focused and attentive, their comprehension may still lack that of an expert. How would an adaptive system learn about a user to the extent that although attentive, their comprehension is not optimal and perhaps recommend material to build a foundation then return later? Most scientists in the field would likely argue that this is the purpose for objective questioning as an assessment. However, these assessments cannot distinguish correctly guessed answers, or misunderstanding in the wording of a question leading to an incorrect answer. Additionally, less fixations and longer saccades may be indicative of proficient comprehension and has been shown to be predictive of higher percentage scores on objective assessments \cite{10.1145/3123024.3123092}.

\section{Conclusion}
In this short review I have discussed ophthalmic neurophysiology and the experimental considerations that must be made to reduce possible noise in an eye-tracking data stream. I have also reviewed the history, experiments, technological benefits, and limitations of eye-tracking studies within the information retrieval field. The concepts of aware and adaptive user interfaces were also explored that humbly motivated my investigations and synthesis of previous work from the fields of industrial engineering, psychophysiology, and information retrieval.

As I stated at the beginning of this review, on the nature of learning I consistently think about my son. Learning from his environment is the foundation of his existence. His interaction with ambient information reinforces or discourages certain behaviors. Throughout this writing I attempted to express these ideas within the context of human-information-machine interaction. More precisely, I attempted to express the need for establishing a foundation that measures the decision making process with lower latency but also with the ability to be operationalized non-intrusively and as an input device, which in order to achieve such a goal, requires a window to the mammalian brain that is achievable only with eye-tracking as I firmly believe this to be the future of ocular navigation for information retrieval.

\bibliographystyle{ACM-Reference-Format}
\bibliography{main.bib}


\begin{thebibliography}{58}


\ifx \showCODEN    \undefined \def \showCODEN     #1{\unskip}     \fi
\ifx \showDOI      \undefined \def \showDOI       #1{#1}\fi
\ifx \showISBNx    \undefined \def \showISBNx     #1{\unskip}     \fi
\ifx \showISBNxiii \undefined \def \showISBNxiii  #1{\unskip}     \fi
\ifx \showISSN     \undefined \def \showISSN      #1{\unskip}     \fi
\ifx \showLCCN     \undefined \def \showLCCN      #1{\unskip}     \fi
\ifx \shownote     \undefined \def \shownote      #1{#1}          \fi
\ifx \showarticletitle \undefined \def \showarticletitle #1{#1}   \fi
\ifx \showURL      \undefined \def \showURL       {\relax}        \fi
\providecommand\bibfield[2]{#2}
\providecommand\bibinfo[2]{#2}
\providecommand\natexlab[1]{#1}
\providecommand\showeprint[2][]{arXiv:#2}

\bibitem[\protect\citeauthoryear{Ahlberg and Shneiderman}{Ahlberg and
  Shneiderman}{2003}]%
        {ahlberg2003visual}
\bibfield{author}{\bibinfo{person}{Christopher Ahlberg} {and}
  \bibinfo{person}{Ben Shneiderman}.} \bibinfo{year}{2003}\natexlab{}.
\newblock \showarticletitle{Visual information seeking: Tight coupling of
  dynamic query filters with starfield displays}.
\newblock In \bibinfo{booktitle}{\emph{The craft of information
  visualization}}. \bibinfo{publisher}{Elsevier}, \bibinfo{pages}{7--13}.
\newblock


\bibitem[\protect\citeauthoryear{Al~Maqbali, Scholer, Thom, and Wu}{Al~Maqbali
  et~al\mbox{.}}{2013}]%
        {10.1145/2537734.2537747}
\bibfield{author}{\bibinfo{person}{Hilal Al~Maqbali}, \bibinfo{person}{Falk
  Scholer}, \bibinfo{person}{James~A. Thom}, {and} \bibinfo{person}{Mingfang
  Wu}.} \bibinfo{year}{2013}\natexlab{}.
\newblock \showarticletitle{Using Eye Tracking for Evaluating Web Search
  Interfaces}. In \bibinfo{booktitle}{\emph{Proceedings of the 18th
  Australasian Document Computing Symposium}} \emph{(\bibinfo{series}{ADCS
  ’13})}. \bibinfo{publisher}{Association for Computing Machinery},
  \bibinfo{address}{New York, NY, USA}, \bibinfo{pages}{2–9}.
\newblock
\showISBNx{9781450325240}
\urldef\tempurl%
\url{https://doi.org/10.1145/2537734.2537747}
\showDOI{\tempurl}


\bibitem[\protect\citeauthoryear{Aula, Majaranta, and R\"{a}ih\"{a}}{Aula
  et~al\mbox{.}}{2005}]%
        {10.1007/11555261_104}
\bibfield{author}{\bibinfo{person}{Anne Aula}, \bibinfo{person}{P\"{a}ivi
  Majaranta}, {and} \bibinfo{person}{Kari-Jouko R\"{a}ih\"{a}}.}
  \bibinfo{year}{2005}\natexlab{}.
\newblock \showarticletitle{Eye-Tracking Reveals the Personal Styles for Search
  Result Evaluation}. In \bibinfo{booktitle}{\emph{Proceedings of the 2005 IFIP
  TC13 International Conference on Human-Computer Interaction}}
  \emph{(\bibinfo{series}{INTERACT’05})}.
  \bibinfo{publisher}{Springer-Verlag}, \bibinfo{address}{Berlin, Heidelberg},
  \bibinfo{pages}{1058–1061}.
\newblock
\showISBNx{3540289437}
\urldef\tempurl%
\url{https://doi.org/10.1007/11555261_104}
\showDOI{\tempurl}


\bibitem[\protect\citeauthoryear{Balatsoukas and Ruthven}{Balatsoukas and
  Ruthven}{2010}]%
        {doi:10.1002/meet.14504701145}
\bibfield{author}{\bibinfo{person}{Panos Balatsoukas} {and}
  \bibinfo{person}{Ian Ruthven}.} \bibinfo{year}{2010}\natexlab{}.
\newblock \showarticletitle{The use of relevance criteria during predictive
  judgment: An eye tracking approach}.
\newblock \bibinfo{journal}{\emph{Proceedings of the American Society for
  Information Science and Technology}} \bibinfo{volume}{47},
  \bibinfo{number}{1} (\bibinfo{year}{2010}), \bibinfo{pages}{1--10}.
\newblock
\urldef\tempurl%
\url{https://doi.org/10.1002/meet.14504701145}
\showDOI{\tempurl}
\showeprint{https://asistdl.onlinelibrary.wiley.com/doi/pdf/10.1002/meet.14504701145}


\bibitem[\protect\citeauthoryear{Balatsoukas and Ruthven}{Balatsoukas and
  Ruthven}{2012}]%
        {10.1002/asi.22707}
\bibfield{author}{\bibinfo{person}{Panos Balatsoukas} {and}
  \bibinfo{person}{Ian Ruthven}.} \bibinfo{year}{2012}\natexlab{}.
\newblock \showarticletitle{An Eye-Tracking Approach to the Analysis of
  Relevance Judgments on the Web: The Case of Google Search Engine}.
\newblock \bibinfo{journal}{\emph{J. Am. Soc. Inf. Sci. Technol.}}
  \bibinfo{volume}{63}, \bibinfo{number}{9} (\bibinfo{date}{Sept.}
  \bibinfo{year}{2012}), \bibinfo{pages}{1728–1746}.
\newblock
\showISSN{1532-2882}
\urldef\tempurl%
\url{https://doi.org/10.1002/asi.22707}
\showDOI{\tempurl}


\bibitem[\protect\citeauthoryear{Bhattacharya, Rakshit, Gwizdka, and
  Kogut}{Bhattacharya et~al\mbox{.}}{2020}]%
        {bhattacharya2020relevance}
\bibfield{author}{\bibinfo{person}{Nilavra Bhattacharya},
  \bibinfo{person}{Somnath Rakshit}, \bibinfo{person}{Jacek Gwizdka}, {and}
  \bibinfo{person}{Paul Kogut}.} \bibinfo{year}{2020}\natexlab{}.
\newblock \showarticletitle{Relevance Prediction from Eye-movements Using
  Semi-interpretable Convolutional Neural Networks}.
\newblock \bibinfo{journal}{\emph{arXiv preprint arXiv:2001.05152}}
  (\bibinfo{year}{2020}).
\newblock


\bibitem[\protect\citeauthoryear{Bilal and Gwizdka}{Bilal and Gwizdka}{2016}]%
        {10.5555/3017447.3017536}
\bibfield{author}{\bibinfo{person}{Dania Bilal} {and} \bibinfo{person}{Jacek
  Gwizdka}.} \bibinfo{year}{2016}\natexlab{}.
\newblock \showarticletitle{Children’s Eye-Fixations on Google Search
  Results}. In \bibinfo{booktitle}{\emph{Proceedings of the 79th ASIS\&T Annual
  Meeting: Creating Knowledge, Enhancing Lives through Information \&
  Technology}} \emph{(\bibinfo{series}{ASIST ’16})}.
  \bibinfo{publisher}{American Society for Information Science},
  \bibinfo{address}{USA}, Article \bibinfo{articleno}{Article 89},
  \bibinfo{numpages}{6}~pages.
\newblock


\bibitem[\protect\citeauthoryear{{Capra}, {Kules}, {Banta}, and
  {Sierra}}{{Capra} et~al\mbox{.}}{2009}]%
        {capra2009faceted}
\bibfield{author}{\bibinfo{person}{Robert {Capra}}, \bibinfo{person}{Bill
  {Kules}}, \bibinfo{person}{Matthew {Banta}}, {and} \bibinfo{person}{Tito
  {Sierra}}.} \bibinfo{year}{2009}\natexlab{}.
\newblock \showarticletitle{Faceted Search for Library Catalogs: Developing
  Grounded Tasks and Analyzing Eye-Tracking Data}.
\newblock


\bibitem[\protect\citeauthoryear{Card}{Card}{1999}]%
        {card1999readings}
\bibfield{author}{\bibinfo{person}{Mackinlay Card}.}
  \bibinfo{year}{1999}\natexlab{}.
\newblock \bibinfo{booktitle}{\emph{Readings in information visualization:
  using vision to think}}.
\newblock \bibinfo{publisher}{Morgan Kaufmann}.
\newblock


\bibitem[\protect\citeauthoryear{Cutrell and Guan}{Cutrell and Guan}{2007}]%
        {10.1145/1240624.1240690}
\bibfield{author}{\bibinfo{person}{Edward Cutrell} {and}
  \bibinfo{person}{Zhiwei Guan}.} \bibinfo{year}{2007}\natexlab{}.
\newblock \showarticletitle{What Are You Looking for? An Eye-Tracking Study of
  Information Usage in Web Search}. In \bibinfo{booktitle}{\emph{Proceedings of
  the SIGCHI Conference on Human Factors in Computing Systems}}
  \emph{(\bibinfo{series}{CHI ’07})}. \bibinfo{publisher}{Association for
  Computing Machinery}, \bibinfo{address}{New York, NY, USA},
  \bibinfo{pages}{407–416}.
\newblock
\showISBNx{9781595935939}
\urldef\tempurl%
\url{https://doi.org/10.1145/1240624.1240690}
\showDOI{\tempurl}


\bibitem[\protect\citeauthoryear{Dinet, Bastien, and Kitajima}{Dinet
  et~al\mbox{.}}{2010}]%
        {10.1145/1941007.1941022}
\bibfield{author}{\bibinfo{person}{J\'{e}r\^{o}me Dinet},
  \bibinfo{person}{J.~M.~Christian Bastien}, {and} \bibinfo{person}{Muneo
  Kitajima}.} \bibinfo{year}{2010}\natexlab{}.
\newblock \showarticletitle{What, Where and How Are Young People Looking for in
  a Search Engine Results Page? Impact of Typographical Cues and Prior Domain
  Knowledge}. In \bibinfo{booktitle}{\emph{Proceedings of the 22nd Conference
  on l’Interaction Homme-Machine}} \emph{(\bibinfo{series}{IHM ’10})}.
  \bibinfo{publisher}{Association for Computing Machinery},
  \bibinfo{address}{New York, NY, USA}, \bibinfo{pages}{105–112}.
\newblock
\showISBNx{9781450304108}
\urldef\tempurl%
\url{https://doi.org/10.1145/1941007.1941022}
\showDOI{\tempurl}


\bibitem[\protect\citeauthoryear{Dumais, Buscher, and Cutrell}{Dumais
  et~al\mbox{.}}{2010}]%
        {10.1145/1840784.1840812}
\bibfield{author}{\bibinfo{person}{Susan~T. Dumais}, \bibinfo{person}{Georg
  Buscher}, {and} \bibinfo{person}{Edward Cutrell}.}
  \bibinfo{year}{2010}\natexlab{}.
\newblock \showarticletitle{Individual Differences in Gaze Patterns for Web
  Search}. In \bibinfo{booktitle}{\emph{Proceedings of the Third Symposium on
  Information Interaction in Context}} \emph{(\bibinfo{series}{IIiX ’10})}.
  \bibinfo{publisher}{Association for Computing Machinery},
  \bibinfo{address}{New York, NY, USA}, \bibinfo{pages}{185–194}.
\newblock
\showISBNx{9781450302470}
\urldef\tempurl%
\url{https://doi.org/10.1145/1840784.1840812}
\showDOI{\tempurl}


\bibitem[\protect\citeauthoryear{Eickhoff, Dungs, and Tran}{Eickhoff
  et~al\mbox{.}}{2015}]%
        {10.1145/2766462.2767703}
\bibfield{author}{\bibinfo{person}{Carsten Eickhoff},
  \bibinfo{person}{Sebastian Dungs}, {and} \bibinfo{person}{Vu Tran}.}
  \bibinfo{year}{2015}\natexlab{}.
\newblock \showarticletitle{An Eye-Tracking Study of Query Reformulation}. In
  \bibinfo{booktitle}{\emph{Proceedings of the 38th International ACM SIGIR
  Conference on Research and Development in Information Retrieval}}
  \emph{(\bibinfo{series}{SIGIR ’15})}. \bibinfo{publisher}{Association for
  Computing Machinery}, \bibinfo{address}{New York, NY, USA},
  \bibinfo{pages}{13–22}.
\newblock
\showISBNx{9781450336215}
\urldef\tempurl%
\url{https://doi.org/10.1145/2766462.2767703}
\showDOI{\tempurl}


\bibitem[\protect\citeauthoryear{Elvesjo, Skogo, and Elvers}{Elvesjo
  et~al\mbox{.}}{2009}]%
        {elvesjo2009method}
\bibfield{author}{\bibinfo{person}{John Elvesjo}, \bibinfo{person}{Marten
  Skogo}, {and} \bibinfo{person}{Gunnar Elvers}.}
  \bibinfo{year}{2009}\natexlab{}.
\newblock \bibinfo{title}{Method and installation for detecting and following
  an eye and the gaze direction thereof}.
\newblock
\newblock
\newblock
\shownote{US Patent 7,572,008.}


\bibitem[\protect\citeauthoryear{Faro, Giordano, Pino, and Spampinato}{Faro
  et~al\mbox{.}}{2010}]%
        {10.1145/1743666.1743684}
\bibfield{author}{\bibinfo{person}{A. Faro}, \bibinfo{person}{D. Giordano},
  \bibinfo{person}{C. Pino}, {and} \bibinfo{person}{C. Spampinato}.}
  \bibinfo{year}{2010}\natexlab{}.
\newblock \showarticletitle{Visual Attention for Implicit Relevance Feedback in
  a Content Based Image Retrieval}. In \bibinfo{booktitle}{\emph{Proceedings of
  the 2010 Symposium on Eye-Tracking Research \& Applications}}
  \emph{(\bibinfo{series}{ETRA ’10})}. \bibinfo{publisher}{Association for
  Computing Machinery}, \bibinfo{address}{New York, NY, USA},
  \bibinfo{pages}{73–76}.
\newblock
\showISBNx{9781605589947}
\urldef\tempurl%
\url{https://doi.org/10.1145/1743666.1743684}
\showDOI{\tempurl}


\bibitem[\protect\citeauthoryear{Ferreira, Apel, and Henderson}{Ferreira
  et~al\mbox{.}}{2008}]%
        {ferreira2008taking}
\bibfield{author}{\bibinfo{person}{Fernanda Ferreira}, \bibinfo{person}{Jens
  Apel}, {and} \bibinfo{person}{John~M Henderson}.}
  \bibinfo{year}{2008}\natexlab{}.
\newblock \showarticletitle{Taking a new look at looking at nothing}.
\newblock \bibinfo{journal}{\emph{Trends in cognitive sciences}}
  \bibinfo{volume}{12}, \bibinfo{number}{11} (\bibinfo{year}{2008}),
  \bibinfo{pages}{405--410}.
\newblock


\bibitem[\protect\citeauthoryear{Gellrich and Kandzia}{Gellrich and
  Kandzia}{2016}]%
        {gellrich2016purkinje}
\bibfield{author}{\bibinfo{person}{MM Gellrich} {and} \bibinfo{person}{C
  Kandzia}.} \bibinfo{year}{2016}\natexlab{}.
\newblock \showarticletitle{Purkinje images in slit lamp videography: Video
  article}.
\newblock \bibinfo{journal}{\emph{Der Ophthalmologe: Zeitschrift der Deutschen
  Ophthalmologischen Gesellschaft}} \bibinfo{volume}{113}, \bibinfo{number}{9}
  (\bibinfo{year}{2016}), \bibinfo{pages}{789--793}.
\newblock


\bibitem[\protect\citeauthoryear{G\"{o}bel, Kiefer, Giannopoulos, Duchowski,
  and Raubal}{G\"{o}bel et~al\mbox{.}}{2018}]%
        {10.1145/3204493.3204544}
\bibfield{author}{\bibinfo{person}{Fabian G\"{o}bel}, \bibinfo{person}{Peter
  Kiefer}, \bibinfo{person}{Ioannis Giannopoulos}, \bibinfo{person}{Andrew~T.
  Duchowski}, {and} \bibinfo{person}{Martin Raubal}.}
  \bibinfo{year}{2018}\natexlab{}.
\newblock \showarticletitle{Improving Map Reading with Gaze-Adaptive Legends}.
  In \bibinfo{booktitle}{\emph{Proceedings of the 2018 ACM Symposium on Eye
  Tracking Research \& Applications}} \emph{(\bibinfo{series}{ETRA ’18})}.
  \bibinfo{publisher}{Association for Computing Machinery},
  \bibinfo{address}{New York, NY, USA}, Article \bibinfo{articleno}{Article
  29}, \bibinfo{numpages}{9}~pages.
\newblock
\showISBNx{9781450357067}
\urldef\tempurl%
\url{https://doi.org/10.1145/3204493.3204544}
\showDOI{\tempurl}


\bibitem[\protect\citeauthoryear{Gossen, H\"{o}bel, and N\"{u}rnberger}{Gossen
  et~al\mbox{.}}{2014}]%
        {10.1145/2556288.2557031}
\bibfield{author}{\bibinfo{person}{Tatiana Gossen}, \bibinfo{person}{Juliane
  H\"{o}bel}, {and} \bibinfo{person}{Andreas N\"{u}rnberger}.}
  \bibinfo{year}{2014}\natexlab{}.
\newblock \showarticletitle{A Comparative Study about Children’s and
  Adults’ Perception of Targeted Web Search Engines}. In
  \bibinfo{booktitle}{\emph{Proceedings of the SIGCHI Conference on Human
  Factors in Computing Systems}} \emph{(\bibinfo{series}{CHI ’14})}.
  \bibinfo{publisher}{Association for Computing Machinery},
  \bibinfo{address}{New York, NY, USA}, \bibinfo{pages}{1821–1824}.
\newblock
\showISBNx{9781450324731}
\urldef\tempurl%
\url{https://doi.org/10.1145/2556288.2557031}
\showDOI{\tempurl}


\bibitem[\protect\citeauthoryear{Granka, Joachims, and Gay}{Granka
  et~al\mbox{.}}{2004}]%
        {10.1145/1008992.1009079}
\bibfield{author}{\bibinfo{person}{Laura~A. Granka}, \bibinfo{person}{Thorsten
  Joachims}, {and} \bibinfo{person}{Geri Gay}.}
  \bibinfo{year}{2004}\natexlab{}.
\newblock \showarticletitle{Eye-Tracking Analysis of User Behavior in WWW
  Search}. In \bibinfo{booktitle}{\emph{Proceedings of the 27th Annual
  International ACM SIGIR Conference on Research and Development in Information
  Retrieval}} \emph{(\bibinfo{series}{SIGIR ’04})}.
  \bibinfo{publisher}{Association for Computing Machinery},
  \bibinfo{address}{New York, NY, USA}, \bibinfo{pages}{478–479}.
\newblock
\showISBNx{1581138814}
\urldef\tempurl%
\url{https://doi.org/10.1145/1008992.1009079}
\showDOI{\tempurl}


\bibitem[\protect\citeauthoryear{Gwizdka}{Gwizdka}{2014}]%
        {10.1145/2578153.2578198}
\bibfield{author}{\bibinfo{person}{Jacek Gwizdka}.}
  \bibinfo{year}{2014}\natexlab{}.
\newblock \showarticletitle{News Stories Relevance Effects on Eye-Movements}.
  In \bibinfo{booktitle}{\emph{Proceedings of the Symposium on Eye Tracking
  Research and Applications}} \emph{(\bibinfo{series}{ETRA ’14})}.
  \bibinfo{publisher}{Association for Computing Machinery},
  \bibinfo{address}{New York, NY, USA}, \bibinfo{pages}{283–286}.
\newblock
\showISBNx{9781450327510}
\urldef\tempurl%
\url{https://doi.org/10.1145/2578153.2578198}
\showDOI{\tempurl}


\bibitem[\protect\citeauthoryear{Gwizdka, Hosseini, Cole, and Wang}{Gwizdka
  et~al\mbox{.}}{2017}]%
        {10.5555/3204593.3204595}
\bibfield{author}{\bibinfo{person}{Jacek Gwizdka}, \bibinfo{person}{Rahilsadat
  Hosseini}, \bibinfo{person}{Michael Cole}, {and} \bibinfo{person}{Shouyi
  Wang}.} \bibinfo{year}{2017}\natexlab{}.
\newblock \showarticletitle{Temporal Dynamics of Eye-Tracking and EEG during
  Reading and Relevance Decisions}.
\newblock \bibinfo{journal}{\emph{J. Assoc. Inf. Sci. Technol.}}
  \bibinfo{volume}{68}, \bibinfo{number}{10} (\bibinfo{date}{Oct.}
  \bibinfo{year}{2017}), \bibinfo{pages}{2299–2312}.
\newblock
\showISSN{2330-1635}


\bibitem[\protect\citeauthoryear{Gwizdka and Zhang}{Gwizdka and Zhang}{2015}]%
        {10.1145/2766462.2767795}
\bibfield{author}{\bibinfo{person}{Jacek Gwizdka} {and}
  \bibinfo{person}{Yinglong Zhang}.} \bibinfo{year}{2015}\natexlab{}.
\newblock \showarticletitle{Differences in Eye-Tracking Measures Between Visits
  and Revisits to Relevant and Irrelevant Web Pages}. In
  \bibinfo{booktitle}{\emph{Proceedings of the 38th International ACM SIGIR
  Conference on Research and Development in Information Retrieval}}
  \emph{(\bibinfo{series}{SIGIR ’15})}. \bibinfo{publisher}{Association for
  Computing Machinery}, \bibinfo{address}{New York, NY, USA},
  \bibinfo{pages}{811–814}.
\newblock
\showISBNx{9781450336215}
\urldef\tempurl%
\url{https://doi.org/10.1145/2766462.2767795}
\showDOI{\tempurl}


\bibitem[\protect\citeauthoryear{Hofmann, Mitra, Radlinski, and
  Shokouhi}{Hofmann et~al\mbox{.}}{2014}]%
        {10.1145/2661829.2661922}
\bibfield{author}{\bibinfo{person}{Kajta Hofmann}, \bibinfo{person}{Bhaskar
  Mitra}, \bibinfo{person}{Filip Radlinski}, {and} \bibinfo{person}{Milad
  Shokouhi}.} \bibinfo{year}{2014}\natexlab{}.
\newblock \showarticletitle{An Eye-Tracking Study of User Interactions with
  Query Auto Completion}. In \bibinfo{booktitle}{\emph{Proceedings of the 23rd
  ACM International Conference on Conference on Information and Knowledge
  Management}} \emph{(\bibinfo{series}{CIKM ’14})}.
  \bibinfo{publisher}{Association for Computing Machinery},
  \bibinfo{address}{New York, NY, USA}, \bibinfo{pages}{549–558}.
\newblock
\showISBNx{9781450325981}
\urldef\tempurl%
\url{https://doi.org/10.1145/2661829.2661922}
\showDOI{\tempurl}


\bibitem[\protect\citeauthoryear{Huey}{Huey}{1898}]%
        {huey1898preliminary}
\bibfield{author}{\bibinfo{person}{Edmund~B Huey}.}
  \bibinfo{year}{1898}\natexlab{}.
\newblock \showarticletitle{Preliminary experiments in the physiology and
  psychology of reading}.
\newblock \bibinfo{journal}{\emph{The American Journal of Psychology}}
  \bibinfo{volume}{9}, \bibinfo{number}{4} (\bibinfo{year}{1898}),
  \bibinfo{pages}{575--586}.
\newblock


\bibitem[\protect\citeauthoryear{Huey}{Huey}{1908}]%
        {huey1908psychology}
\bibfield{author}{\bibinfo{person}{Edmund~Burke Huey}.}
  \bibinfo{year}{1908}\natexlab{}.
\newblock \bibinfo{booktitle}{\emph{The psychology and pedagogy of reading}}.
\newblock \bibinfo{publisher}{The Macmillan Company}.
\newblock


\bibitem[\protect\citeauthoryear{Hyrskykari, Majaranta, Aaltonen, and
  R\"{a}ih\"{a}}{Hyrskykari et~al\mbox{.}}{2000}]%
        {10.1145/355017.355019}
\bibfield{author}{\bibinfo{person}{Aulikki Hyrskykari},
  \bibinfo{person}{P\"{a}ivi Majaranta}, \bibinfo{person}{Antti Aaltonen},
  {and} \bibinfo{person}{Kari-Jouko R\"{a}ih\"{a}}.}
  \bibinfo{year}{2000}\natexlab{}.
\newblock \showarticletitle{Design Issues of IDICT: A Gaze-Assisted Translation
  Aid}. In \bibinfo{booktitle}{\emph{Proceedings of the 2000 Symposium on Eye
  Tracking Research \& Applications}} \emph{(\bibinfo{series}{ETRA ’00})}.
  \bibinfo{publisher}{Association for Computing Machinery},
  \bibinfo{address}{New York, NY, USA}, \bibinfo{pages}{9–14}.
\newblock
\showISBNx{1581132808}
\urldef\tempurl%
\url{https://doi.org/10.1145/355017.355019}
\showDOI{\tempurl}


\bibitem[\protect\citeauthoryear{Ishita, Mine, Koizumi, Miyata, Kunimoto,
  Shiozaki, Kurata, and Ueda}{Ishita et~al\mbox{.}}{2009}]%
        {10.1145/1555400.1555485}
\bibfield{author}{\bibinfo{person}{Emi Ishita}, \bibinfo{person}{Shinji Mine},
  \bibinfo{person}{Masanori Koizumi}, \bibinfo{person}{Yosuke Miyata},
  \bibinfo{person}{Chihiro Kunimoto}, \bibinfo{person}{Junko Shiozaki},
  \bibinfo{person}{Keiko Kurata}, {and} \bibinfo{person}{Shuichi Ueda}.}
  \bibinfo{year}{2009}\natexlab{}.
\newblock \showarticletitle{Analyzing OPAC Use with Screen Views and Eye
  Tracking}. In \bibinfo{booktitle}{\emph{Proceedings of the 9th ACM/IEEE-CS
  Joint Conference on Digital Libraries}} \emph{(\bibinfo{series}{JCDL
  ’09})}. \bibinfo{publisher}{Association for Computing Machinery},
  \bibinfo{address}{New York, NY, USA}, \bibinfo{pages}{405–406}.
\newblock
\showISBNx{9781605583228}
\urldef\tempurl%
\url{https://doi.org/10.1145/1555400.1555485}
\showDOI{\tempurl}


\bibitem[\protect\citeauthoryear{Jacucci, Barral, Daee, Wenzel, Serim,
  Ruotsalo, Pluchino, Freeman, Gamberini, Kaski, and Blankertz}{Jacucci
  et~al\mbox{.}}{2019}]%
        {doi:10.1002/asi.24161}
\bibfield{author}{\bibinfo{person}{Giulio Jacucci}, \bibinfo{person}{Oswald
  Barral}, \bibinfo{person}{Pedram Daee}, \bibinfo{person}{Markus Wenzel},
  \bibinfo{person}{Baris Serim}, \bibinfo{person}{Tuukka Ruotsalo},
  \bibinfo{person}{Patrik Pluchino}, \bibinfo{person}{Jonathan Freeman},
  \bibinfo{person}{Luciano Gamberini}, \bibinfo{person}{Samuel Kaski}, {and}
  \bibinfo{person}{Benjamin Blankertz}.} \bibinfo{year}{2019}\natexlab{}.
\newblock \showarticletitle{Integrating neurophysiologic relevance feedback in
  intent modeling for information retrieval}.
\newblock \bibinfo{journal}{\emph{Journal of the Association for Information
  Science and Technology}} \bibinfo{volume}{70}, \bibinfo{number}{9}
  (\bibinfo{year}{2019}), \bibinfo{pages}{917--930}.
\newblock
\urldef\tempurl%
\url{https://doi.org/10.1002/asi.24161}
\showDOI{\tempurl}
\showeprint{https://asistdl.onlinelibrary.wiley.com/doi/pdf/10.1002/asi.24161}


\bibitem[\protect\citeauthoryear{Javal}{Javal}{1878}]%
        {javal1878essai}
\bibfield{author}{\bibinfo{person}{Emile Javal}.}
  \bibinfo{year}{1878}\natexlab{}.
\newblock \showarticletitle{Essai sur la physiologie de la lecture}.
\newblock \bibinfo{journal}{\emph{Annales d'Ocilistique}}  \bibinfo{volume}{80}
  (\bibinfo{year}{1878}), \bibinfo{pages}{97--117}.
\newblock


\bibitem[\protect\citeauthoryear{Kl\"{o}ckner, Wirschum, and
  Jameson}{Kl\"{o}ckner et~al\mbox{.}}{2004}]%
        {10.1145/985921.986115}
\bibfield{author}{\bibinfo{person}{Kerstin Kl\"{o}ckner},
  \bibinfo{person}{Nadine Wirschum}, {and} \bibinfo{person}{Anthony Jameson}.}
  \bibinfo{year}{2004}\natexlab{}.
\newblock \showarticletitle{Depth- and Breadth-First Processing of Search
  Result Lists}. In \bibinfo{booktitle}{\emph{CHI ’04 Extended Abstracts on
  Human Factors in Computing Systems}} \emph{(\bibinfo{series}{CHI EA ’04})}.
  \bibinfo{publisher}{Association for Computing Machinery},
  \bibinfo{address}{New York, NY, USA}, \bibinfo{pages}{1539}.
\newblock
\showISBNx{1581137036}
\urldef\tempurl%
\url{https://doi.org/10.1145/985921.986115}
\showDOI{\tempurl}


\bibitem[\protect\citeauthoryear{Landolt}{Landolt}{1879}]%
        {landolt1879manual}
\bibfield{author}{\bibinfo{person}{Edmond Landolt}.}
  \bibinfo{year}{1879}\natexlab{}.
\newblock \bibinfo{booktitle}{\emph{A Manual of Examination of the Eyes...}}
\newblock \bibinfo{publisher}{Bailli{\`e}re, Tindall \& Cox}.
\newblock


\bibitem[\protect\citeauthoryear{Liu, Liu, Zhou, Zhang, and Ma}{Liu
  et~al\mbox{.}}{2015}]%
        {10.1145/2766462.2767714}
\bibfield{author}{\bibinfo{person}{Zeyang Liu}, \bibinfo{person}{Yiqun Liu},
  \bibinfo{person}{Ke Zhou}, \bibinfo{person}{Min Zhang}, {and}
  \bibinfo{person}{Shaoping Ma}.} \bibinfo{year}{2015}\natexlab{}.
\newblock \showarticletitle{Influence of Vertical Result in Web Search
  Examination}. In \bibinfo{booktitle}{\emph{Proceedings of the 38th
  International ACM SIGIR Conference on Research and Development in Information
  Retrieval}} \emph{(\bibinfo{series}{SIGIR ’15})}.
  \bibinfo{publisher}{Association for Computing Machinery},
  \bibinfo{address}{New York, NY, USA}, \bibinfo{pages}{193–202}.
\newblock
\showISBNx{9781450336215}
\urldef\tempurl%
\url{https://doi.org/10.1145/2766462.2767714}
\showDOI{\tempurl}


\bibitem[\protect\citeauthoryear{Low, Bubalo, Gossen, Kotzyba, Brechmann,
  Huckauf, and N\"{u}rnberger}{Low et~al\mbox{.}}{2017}]%
        {10.1145/3020165.3022131}
\bibfield{author}{\bibinfo{person}{Thomas Low}, \bibinfo{person}{Nikola
  Bubalo}, \bibinfo{person}{Tatiana Gossen}, \bibinfo{person}{Michael Kotzyba},
  \bibinfo{person}{Andr\'{e} Brechmann}, \bibinfo{person}{Anke Huckauf}, {and}
  \bibinfo{person}{Andreas N\"{u}rnberger}.} \bibinfo{year}{2017}\natexlab{}.
\newblock \showarticletitle{Towards Identifying User Intentions in Exploratory
  Search Using Gaze and Pupil Tracking}. In
  \bibinfo{booktitle}{\emph{Proceedings of the 2017 Conference on Conference
  Human Information Interaction and Retrieval}} \emph{(\bibinfo{series}{CHIIR
  ’17})}. \bibinfo{publisher}{Association for Computing Machinery},
  \bibinfo{address}{New York, NY, USA}, \bibinfo{pages}{273–276}.
\newblock
\showISBNx{9781450346771}
\urldef\tempurl%
\url{https://doi.org/10.1145/3020165.3022131}
\showDOI{\tempurl}


\bibitem[\protect\citeauthoryear{Ludwig and Czyz}{Ludwig and Czyz}{2019}]%
        {ludwig2019anatomy}
\bibfield{author}{\bibinfo{person}{Parker~E Ludwig} {and}
  \bibinfo{person}{Craig~N Czyz}.} \bibinfo{year}{2019}\natexlab{}.
\newblock \showarticletitle{Anatomy, Head and Neck, Eye Muscles}.
\newblock In \bibinfo{booktitle}{\emph{StatPearls [Internet]}}.
  \bibinfo{publisher}{StatPearls Publishing}.
\newblock


\bibitem[\protect\citeauthoryear{Marcos, Nettleton, and
  S\'{a}ez-Trumper}{Marcos et~al\mbox{.}}{2012}]%
        {10.5555/2377916.2377949}
\bibfield{author}{\bibinfo{person}{Mari-Carmen Marcos},
  \bibinfo{person}{David~F. Nettleton}, {and} \bibinfo{person}{Diego
  S\'{a}ez-Trumper}.} \bibinfo{year}{2012}\natexlab{}.
\newblock \showarticletitle{A User Study of Web Search Session Behaviour Using
  Eye Tracking Data}. In \bibinfo{booktitle}{\emph{Proceedings of the 26th
  Annual BCS Interaction Specialist Group Conference on People and Computers}}
  \emph{(\bibinfo{series}{BCS-HCI ’12})}. \bibinfo{publisher}{BCS Learning \&
  Development Ltd.}, \bibinfo{address}{Swindon, GBR},
  \bibinfo{pages}{262–267}.
\newblock


\bibitem[\protect\citeauthoryear{Merchant, Morrissette, and
  Porterfield}{Merchant et~al\mbox{.}}{1974}]%
        {merchant1974remote}
\bibfield{author}{\bibinfo{person}{John Merchant}, \bibinfo{person}{Richard
  Morrissette}, {and} \bibinfo{person}{James~L Porterfield}.}
  \bibinfo{year}{1974}\natexlab{}.
\newblock \showarticletitle{Remote measurement of eye direction allowing
  subject motion over one cubic foot of space}.
\newblock \bibinfo{journal}{\emph{IEEE transactions on biomedical engineering}}
  \bibinfo{number}{4} (\bibinfo{year}{1974}), \bibinfo{pages}{309--317}.
\newblock


\bibitem[\protect\citeauthoryear{Mostafa and Gwizdka}{Mostafa and
  Gwizdka}{2016}]%
        {Mostafa2016DeepeningTR}
\bibfield{author}{\bibinfo{person}{Javed Mostafa} {and} \bibinfo{person}{Jacek
  Gwizdka}.} \bibinfo{year}{2016}\natexlab{}.
\newblock \showarticletitle{Deepening the Role of the User: Neuro-Physiological
  Evidence as a Basis for Studying and Improving Search}. In
  \bibinfo{booktitle}{\emph{CHIIR '16}}.
\newblock


\bibitem[\protect\citeauthoryear{Murata and Moriwaka}{Murata and
  Moriwaka}{2009}]%
        {murata2009basic}
\bibfield{author}{\bibinfo{person}{Atsuo Murata} {and} \bibinfo{person}{Makoto
  Moriwaka}.} \bibinfo{year}{2009}\natexlab{}.
\newblock \showarticletitle{Basic study for development of web browser suitable
  for eye-gaze input system-identification of optimal click method}. In
  \bibinfo{booktitle}{\emph{Proceedings: Fifth International Workshop on
  Computational Intelligence \& Applications}}, Vol.~\bibinfo{volume}{2009}.
  IEEE SMC Hiroshima Chapter, \bibinfo{pages}{302--305}.
\newblock


\bibitem[\protect\citeauthoryear{Nguyen and Liu}{Nguyen and Liu}{2016}]%
        {10.1145/2858036.2858137}
\bibfield{author}{\bibinfo{person}{Cuong Nguyen} {and} \bibinfo{person}{Feng
  Liu}.} \bibinfo{year}{2016}\natexlab{}.
\newblock \showarticletitle{Gaze-Based Notetaking for Learning from Lecture
  Videos}. In \bibinfo{booktitle}{\emph{Proceedings of the 2016 CHI Conference
  on Human Factors in Computing Systems}} \emph{(\bibinfo{series}{CHI ’16})}.
  \bibinfo{publisher}{Association for Computing Machinery},
  \bibinfo{address}{New York, NY, USA}, \bibinfo{pages}{2093–2097}.
\newblock
\showISBNx{9781450333627}
\urldef\tempurl%
\url{https://doi.org/10.1145/2858036.2858137}
\showDOI{\tempurl}


\bibitem[\protect\citeauthoryear{Nikolov, Newman, Bull, Canagarajah, Jones, and
  Gilchrist}{Nikolov et~al\mbox{.}}{2004}]%
        {10.1145/968363.968366}
\bibfield{author}{\bibinfo{person}{Stavri~G. Nikolov},
  \bibinfo{person}{Timothy~D. Newman}, \bibinfo{person}{Dave~R. Bull},
  \bibinfo{person}{Nishan~C. Canagarajah}, \bibinfo{person}{Michael~G. Jones},
  {and} \bibinfo{person}{Iain~D. Gilchrist}.} \bibinfo{year}{2004}\natexlab{}.
\newblock \showarticletitle{Gaze-Contingent Display Using Texture Mapping and
  OpenGL: System and Applications}. In \bibinfo{booktitle}{\emph{Proceedings of
  the 2004 Symposium on Eye Tracking Research \& Applications}}
  \emph{(\bibinfo{series}{ETRA ’04})}. \bibinfo{publisher}{Association for
  Computing Machinery}, \bibinfo{address}{New York, NY, USA},
  \bibinfo{pages}{11–18}.
\newblock
\showISBNx{1581138253}
\urldef\tempurl%
\url{https://doi.org/10.1145/968363.968366}
\showDOI{\tempurl}


\bibitem[\protect\citeauthoryear{Oyekoya and Stentiford}{Oyekoya and
  Stentiford}{2006}]%
        {10.1145/1117309.1117324}
\bibfield{author}{\bibinfo{person}{Oyewole Oyekoya} {and} \bibinfo{person}{Fred
  Stentiford}.} \bibinfo{year}{2006}\natexlab{}.
\newblock \showarticletitle{An Eye Tracking Interface for Image Search}. In
  \bibinfo{booktitle}{\emph{Proceedings of the 2006 Symposium on Eye Tracking
  Research \& Applications}} \emph{(\bibinfo{series}{ETRA ’06})}.
  \bibinfo{publisher}{Association for Computing Machinery},
  \bibinfo{address}{New York, NY, USA}, \bibinfo{pages}{40}.
\newblock
\showISBNx{1595933050}
\urldef\tempurl%
\url{https://doi.org/10.1145/1117309.1117324}
\showDOI{\tempurl}


\bibitem[\protect\citeauthoryear{Palinko, Rea, Sandini, and Sciutti}{Palinko
  et~al\mbox{.}}{2016}]%
        {10.1145/2857491.2888590}
\bibfield{author}{\bibinfo{person}{Oskar Palinko}, \bibinfo{person}{Francesco
  Rea}, \bibinfo{person}{Giulio Sandini}, {and} \bibinfo{person}{Alessandra
  Sciutti}.} \bibinfo{year}{2016}\natexlab{}.
\newblock \showarticletitle{Eye Tracking for Human Robot Interaction}. In
  \bibinfo{booktitle}{\emph{Proceedings of the Ninth Biennial ACM Symposium on
  Eye Tracking Research \& Applications}} \emph{(\bibinfo{series}{ETRA
  ’16})}. \bibinfo{publisher}{Association for Computing Machinery},
  \bibinfo{address}{New York, NY, USA}, \bibinfo{pages}{327–328}.
\newblock
\showISBNx{9781450341257}
\urldef\tempurl%
\url{https://doi.org/10.1145/2857491.2888590}
\showDOI{\tempurl}


\bibitem[\protect\citeauthoryear{Perlman, Kolb, and Nelson}{Perlman
  et~al\mbox{.}}{2015}]%
        {perlman2015organization}
\bibfield{author}{\bibinfo{person}{Ido Perlman}, \bibinfo{person}{Helga Kolb},
  {and} \bibinfo{person}{Ralph Nelson}.} \bibinfo{year}{2015}\natexlab{}.
\newblock \bibinfo{title}{The Organization of the Retina and Visual System}.
\newblock
\newblock


\bibitem[\protect\citeauthoryear{Pirolli, Card, and Van Der~Wege}{Pirolli
  et~al\mbox{.}}{2001}]%
        {pirolli2001visual}
\bibfield{author}{\bibinfo{person}{Peter Pirolli}, \bibinfo{person}{Stuart~K
  Card}, {and} \bibinfo{person}{Mija~M Van Der~Wege}.}
  \bibinfo{year}{2001}\natexlab{}.
\newblock \showarticletitle{Visual information foraging in a focus+ context
  visualization}. In \bibinfo{booktitle}{\emph{Proceedings of the SIGCHI
  conference on Human factors in computing systems}}.
  \bibinfo{pages}{506--513}.
\newblock


\bibitem[\protect\citeauthoryear{Pirolli, Card, and Van Der~Wege}{Pirolli
  et~al\mbox{.}}{2003}]%
        {pirolli2003effects}
\bibfield{author}{\bibinfo{person}{Peter Pirolli}, \bibinfo{person}{Stuart~K
  Card}, {and} \bibinfo{person}{Mija~M Van Der~Wege}.}
  \bibinfo{year}{2003}\natexlab{}.
\newblock \showarticletitle{The effects of information scent on visual search
  in the hyperbolic tree browser}.
\newblock \bibinfo{journal}{\emph{ACM Transactions on Computer-Human
  Interaction (TOCHI)}} \bibinfo{volume}{10}, \bibinfo{number}{1}
  (\bibinfo{year}{2003}), \bibinfo{pages}{20--53}.
\newblock


\bibitem[\protect\citeauthoryear{Potter, Wyble, Hagmann, and McCourt}{Potter
  et~al\mbox{.}}{2014}]%
        {potter2014detecting}
\bibfield{author}{\bibinfo{person}{Mary~C Potter}, \bibinfo{person}{Brad
  Wyble}, \bibinfo{person}{Carl~Erick Hagmann}, {and} \bibinfo{person}{Emily~S
  McCourt}.} \bibinfo{year}{2014}\natexlab{}.
\newblock \showarticletitle{Detecting meaning in RSVP at 13 ms per picture}.
\newblock \bibinfo{journal}{\emph{Attention, Perception, \& Psychophysics}}
  \bibinfo{volume}{76}, \bibinfo{number}{2} (\bibinfo{year}{2014}),
  \bibinfo{pages}{270--279}.
\newblock


\bibitem[\protect\citeauthoryear{Ramloll, Trepagnier, Sebrechts, and
  Finkelmeyer}{Ramloll et~al\mbox{.}}{2004}]%
        {10.1145/968363.968367}
\bibfield{author}{\bibinfo{person}{Rameshsharma Ramloll},
  \bibinfo{person}{Cheryl Trepagnier}, \bibinfo{person}{Marc Sebrechts}, {and}
  \bibinfo{person}{Andreas Finkelmeyer}.} \bibinfo{year}{2004}\natexlab{}.
\newblock \showarticletitle{A Gaze Contingent Environment for Fostering Social
  Attention in Autistic Children}. In \bibinfo{booktitle}{\emph{Proceedings of
  the 2004 Symposium on Eye Tracking Research \& Applications}}
  \emph{(\bibinfo{series}{ETRA ’04})}. \bibinfo{publisher}{Association for
  Computing Machinery}, \bibinfo{address}{New York, NY, USA},
  \bibinfo{pages}{19–26}.
\newblock
\showISBNx{1581138253}
\urldef\tempurl%
\url{https://doi.org/10.1145/968363.968367}
\showDOI{\tempurl}


\bibitem[\protect\citeauthoryear{Salojarvi, Kojo, Simola, and Kaski}{Salojarvi
  et~al\mbox{.}}{2003}]%
        {SalojarviJarkko}
\bibfield{author}{\bibinfo{person}{Jarkko Salojarvi}, \bibinfo{person}{Ilpo
  Kojo}, \bibinfo{person}{Jaana Simola}, {and} \bibinfo{person}{Samuel Kaski}.}
  \bibinfo{year}{2003}\natexlab{}.
\newblock \showarticletitle{Can relevance be inferred from eye movements in
  information retrieval}.
\newblock \bibinfo{journal}{\emph{Proceedings of WSOM}}  \bibinfo{volume}{3}
  (\bibinfo{date}{01} \bibinfo{year}{2003}).
\newblock


\bibitem[\protect\citeauthoryear{Sanches, Kise, and Augereau}{Sanches
  et~al\mbox{.}}{2017}]%
        {10.1145/3123024.3123092}
\bibfield{author}{\bibinfo{person}{Charles~Lima Sanches},
  \bibinfo{person}{Koichi Kise}, {and} \bibinfo{person}{Olivier Augereau}.}
  \bibinfo{year}{2017}\natexlab{}.
\newblock \showarticletitle{Japanese Reading Objective Understanding Estimation
  by Eye Gaze Analysis}. In \bibinfo{booktitle}{\emph{Proceedings of the 2017
  ACM International Joint Conference on Pervasive and Ubiquitous Computing and
  Proceedings of the 2017 ACM International Symposium on Wearable Computers}}
  \emph{(\bibinfo{series}{UbiComp ’17})}. \bibinfo{publisher}{Association for
  Computing Machinery}, \bibinfo{address}{New York, NY, USA},
  \bibinfo{pages}{121–124}.
\newblock
\showISBNx{9781450351904}
\urldef\tempurl%
\url{https://doi.org/10.1145/3123024.3123092}
\showDOI{\tempurl}


\bibitem[\protect\citeauthoryear{Shackel}{Shackel}{1960a}]%
        {shackel1960note}
\bibfield{author}{\bibinfo{person}{B Shackel}.}
  \bibinfo{year}{1960}\natexlab{a}.
\newblock \showarticletitle{Note on mobile eye viewpoint recording}.
\newblock \bibinfo{journal}{\emph{JOSA}} \bibinfo{volume}{50},
  \bibinfo{number}{8} (\bibinfo{year}{1960}), \bibinfo{pages}{763--768}.
\newblock


\bibitem[\protect\citeauthoryear{Shackel}{Shackel}{1960b}]%
        {shackel1960pilot}
\bibfield{author}{\bibinfo{person}{B Shackel}.}
  \bibinfo{year}{1960}\natexlab{b}.
\newblock \showarticletitle{Pilot study in electro-oculography}.
\newblock \bibinfo{journal}{\emph{The British journal of ophthalmology}}
  \bibinfo{volume}{44}, \bibinfo{number}{2} (\bibinfo{year}{1960}),
  \bibinfo{pages}{89}.
\newblock


\bibitem[\protect\citeauthoryear{Shell, Vertegaal, Cheng, Skaburskis, Sohn,
  Stewart, Aoudeh, and Dickie}{Shell et~al\mbox{.}}{2004}]%
        {10.1145/968363.968384}
\bibfield{author}{\bibinfo{person}{Jeffrey~S. Shell}, \bibinfo{person}{Roel
  Vertegaal}, \bibinfo{person}{Daniel Cheng}, \bibinfo{person}{Alexander~W.
  Skaburskis}, \bibinfo{person}{Changuk Sohn}, \bibinfo{person}{A.~James
  Stewart}, \bibinfo{person}{Omar Aoudeh}, {and} \bibinfo{person}{Connor
  Dickie}.} \bibinfo{year}{2004}\natexlab{}.
\newblock \showarticletitle{ECSGlasses and EyePliances: Using Attention to Open
  Sociable Windows of Interaction}. In \bibinfo{booktitle}{\emph{Proceedings of
  the 2004 Symposium on Eye Tracking Research \& Applications}}
  \emph{(\bibinfo{series}{ETRA ’04})}. \bibinfo{publisher}{Association for
  Computing Machinery}, \bibinfo{address}{New York, NY, USA},
  \bibinfo{pages}{93–100}.
\newblock
\showISBNx{1581138253}
\urldef\tempurl%
\url{https://doi.org/10.1145/968363.968384}
\showDOI{\tempurl}


\bibitem[\protect\citeauthoryear{Shepherd, Findlay, and Hockey}{Shepherd
  et~al\mbox{.}}{1986}]%
        {shepherd1986relationship}
\bibfield{author}{\bibinfo{person}{Martin Shepherd}, \bibinfo{person}{John~M
  Findlay}, {and} \bibinfo{person}{Robert~J Hockey}.}
  \bibinfo{year}{1986}\natexlab{}.
\newblock \showarticletitle{The relationship between eye movements and spatial
  attention}.
\newblock \bibinfo{journal}{\emph{The Quarterly Journal of Experimental
  Psychology}} \bibinfo{volume}{38}, \bibinfo{number}{3}
  (\bibinfo{year}{1986}), \bibinfo{pages}{475--491}.
\newblock


\bibitem[\protect\citeauthoryear{Theeuwes, Belopolsky, and Olivers}{Theeuwes
  et~al\mbox{.}}{2009}]%
        {theeuwes2009interactions}
\bibfield{author}{\bibinfo{person}{Jan Theeuwes}, \bibinfo{person}{Artem
  Belopolsky}, {and} \bibinfo{person}{Christian~NL Olivers}.}
  \bibinfo{year}{2009}\natexlab{}.
\newblock \showarticletitle{Interactions between working memory, attention and
  eye movements}.
\newblock \bibinfo{journal}{\emph{Acta psychologica}} \bibinfo{volume}{132},
  \bibinfo{number}{2} (\bibinfo{year}{2009}), \bibinfo{pages}{106--114}.
\newblock


\bibitem[\protect\citeauthoryear{Wade}{Wade}{2010}]%
        {wade2010pioneers}
\bibfield{author}{\bibinfo{person}{Nicholas~J Wade}.}
  \bibinfo{year}{2010}\natexlab{}.
\newblock \showarticletitle{Pioneers of eye movement research}.
\newblock \bibinfo{journal}{\emph{i-Perception}} \bibinfo{volume}{1},
  \bibinfo{number}{2} (\bibinfo{year}{2010}), \bibinfo{pages}{33--68}.
\newblock


\bibitem[\protect\citeauthoryear{Wu, Liu, Tsai, and Yau}{Wu
  et~al\mbox{.}}{2019}]%
        {doi:10.1002/asi.24240}
\bibfield{author}{\bibinfo{person}{Yingying Wu}, \bibinfo{person}{Yiqun Liu},
  \bibinfo{person}{Yen-Hsi~Richard Tsai}, {and} \bibinfo{person}{Shing-Tung
  Yau}.} \bibinfo{year}{2019}\natexlab{}.
\newblock \showarticletitle{Investigating the role of eye movements and
  physiological signals in search satisfaction prediction using geometric
  analysis}.
\newblock \bibinfo{journal}{\emph{Journal of the Association for Information
  Science and Technology}} \bibinfo{volume}{70}, \bibinfo{number}{9}
  (\bibinfo{year}{2019}), \bibinfo{pages}{981--999}.
\newblock
\urldef\tempurl%
\url{https://doi.org/10.1002/asi.24240}
\showDOI{\tempurl}
\showeprint{https://asistdl.onlinelibrary.wiley.com/doi/pdf/10.1002/asi.24240}


\bibitem[\protect\citeauthoryear{Yarbus}{Yarbus}{[n.d.]}]%
        {yarbuseye}
\bibfield{author}{\bibinfo{person}{Alfred~L Yarbus}.}
  \bibinfo{year}{[n.d.]}\natexlab{}.
\newblock \bibinfo{booktitle}{\emph{Eye Movements and Vision}}.
\newblock \bibinfo{publisher}{Springer}.
\newblock


\end{thebibliography}

\end{document}